\documentclass[journal]{IEEEtran}

\usepackage[ruled,vlined, linesnumbered]{algorithm2e}

\usepackage[flushleft]{threeparttable}
\usepackage{pgfplots}
\usepackage{tikz}
\pgfplotsset{compat=newest}
\usetikzlibrary{plotmarks}
\usetikzlibrary{arrows.meta}
\usepackage{amssymb}
\usepgfplotslibrary{patchplots}

\usepackage[eulergreek]{sansmath}
\usepackage{adjustbox}
\usepackage{enumitem}

\AtBeginDocument{%
  \setlist[enumerate]{%
    leftmargin=1em,
    labelwidth=0.8em,
    labelsep=0.2em,
    align=left,
    itemindent=0pt,
    listparindent=0pt
  }%
}
\usepackage{dirtytalk}

\usepackage[subpreambles=true]{standalone}
\usepackage{import}
\usepackage{comment}
\usepackage[bottom]{footmisc}
\usepackage{amsthm}

\usepackage{amsmath}
\usepackage{xcolor}
\usepackage{amsfonts}
\usepackage{hyperref}
\usepackage{graphicx, float}
\usepackage{subfig}
\usepackage{multirow}
\usepackage{siunitx}
\usepackage{url}
\usepackage{framed}
\usepackage{xfrac}
\usepackage{svg}
\usepackage{breqn}
\usepackage{makecell}
\usepackage{booktabs}
\usepackage{listings}

\usepackage[table]{xcolor}
\usepackage{rotating}
\usepackage{mathtools} \usepackage{xspace} \newcommand{\FLIP}{\protect\reflectbox{F}LIP\xspace}

\usepackage{array}
\newcolumntype{L}[1]{>{\raggedright\let\newline\\\arraybackslash\hspace{0pt}}m{#1}}
\newcolumntype{C}[1]{>{\centering\let\newline\\\arraybackslash\hspace{0pt}}m{#1}}
\newcolumntype{R}[1]{>{\raggedleft\let\newline\\\arraybackslash\hspace{0pt}}m{#1}}
\usepackage{atbegshi}
\AtBeginDocument{\AtBeginShipoutNext{\AtBeginShipoutDiscard}}

\usepackage{etoolbox}
\usepackage{ragged2e}

\newcommand{\insertfig}{%
  \begin{minipage}{\textwidth}
    \centering
    \includegraphics[
      width=\textwidth
    ]{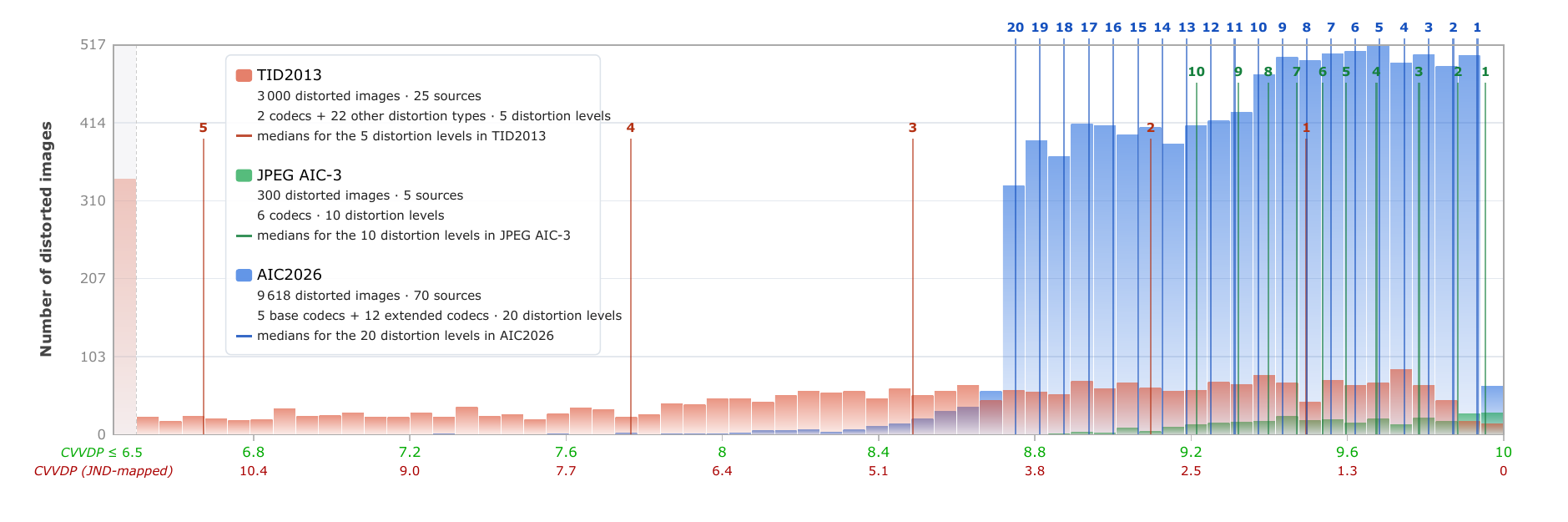}

    \vspace{-12pt}

    \small
    \justifying
    \setlength{\parindent}{0pt}%
    \hangindent=.1em
    \hangafter=1
    \noindent
    Fig.~1: Distributions of ColorVideoVDP (CVVDP) scores for distorted images in the TID2013, JPEG AIC-3, and proposed AIC2026 datasets.
    AIC2026 is a large-scale dataset with diverse image content, fine-grained distortion levels, and a wide range of compression artifacts produced by conventional and learning-based codecs.
    For each dataset, the vertical lines indicate the median CVVDP score at each distortion level.
    \par
  \end{minipage}%
  \vspace{-17pt}
}

\makeatletter
\apptocmd{\@maketitle}{%
  \par
  \insertfig
  \par
}{}{}
\makeatother

\DeclareMathOperator{\imd}{\textsc{imd}}

\begin{document}
\setcounter{page}{0}
\setcounter{figure}{1}
\newcommand{\historef}{1} 

\title{JPEG AIC2026: A large-scale dataset for fine-grained assessment of image coding}

\author{Mohsen Jenadeleh$^a$,
Jon Sneyers$^b$,
João Ascenso$^c$,
Thomas Richter$^d$,
Alexander Karabutov$^e$,
Panqi Jia$^e$,
Elena Alshina$^e$,
Osamu Watanabe$^f$,
António Pinheiro$^g$,
Touradj Ebrahimi$^h$,
Dietmar Saupe$^a$}

\thanks{Funded by the Deutsche Forschungsgemeinschaft (DFG) -- Project ID 496858717  ``Fine-grained visual quality assessment and modeling for high-fidelity compressed images'' and Project ID 251654672 - SFB/Transregio 161.
}

\thanks{$^a$University of Konstanz, Germany.
$^b$Cloudinary, Belgium.
$^c$IST-IT, Portugal.
$^d$Fraunhofer IIS, Germany.
$^e$Huawei, Germany.
$^f$Takushoku University, Japan.
$^g$IT-UBI, Portugal.
$^h$EPFL, Switzerland.
}

\setcounter{page}{0}

\maketitle

\begin{abstract}
Recent advances in conventional and learning-based image coding have increased the demand for benchmark datasets that support fine-grained assessment of compressed image quality, particularly for learning-based image compression methods. This paper introduces Assessment of Image Coding 2026 (AIC2026), a large-scale dataset for high-fidelity image compression containing 70 source images selected from 2,787 candidates using semantic clustering, inter-metric disagreement among objective image quality assessment (IQA) methods, and manual inspection and refinement. The dataset covers a wide range of compression artifacts produced by eight conventional and four learning-based codecs across 17 coding configurations. Each source image is encoded using seven codecs. For each source–codec pair, decoded images are provided at 20 perceptually spaced distortion levels, corresponding approximately to 0.2–4.0 just-noticeable difference (JND) units  using the ColorVideoVDP (CVVDP) metric for distortion estimation, yielding 9,618 distorted images. This fine-grained sampling enables analysis of distortion-rate behavior and objective metric evaluation for subtle quality differences across a wide range of compression artifacts. We report an extensive objective analysis using 24 conventional and 12 learning-based IQA methods. The results show substantial disagreement among current IQA methods for fine-grained quality differences, particularly for artifacts introduced by learning-based codecs.
The complete dataset is publicly available\footnote{Dataset DOI: \url{https://doi.org/10.18419/DARUS-6156}} (CC BY-SA 4.0).
\end{abstract}


\IEEEpeerreviewmaketitle

\section{Introduction}
\label{sec:introduction}
\IEEEPARstart{L}{ossy compression} remains a key area in image processing, due to the growing demand for efficient storage and transmission of visual data. Conventional image codecs such as JPEG~\cite{jpeg}, JPEG~2000~\cite{rabbani2002overview},  HEVC Intra~\cite{sullivan2012overview}, AVIF~\cite{barman2020evaluation}, VVC Intra~\cite{hamidouche2022versatile}, and JPEG~XL~\cite{jxl2025} reduce data redundancy and bitrate using frequency transforms, including the discrete cosine transform (DCT) or the discrete wavelet transform (DWT), and quantization, prediction, and filtering.  
These codecs may introduce visible distortions in the decoded image as the bitrate decreases, including blocking, ringing, blurring, texture distortion, and changes in color or contrast.

Recently, learning-based image compression methods, such as JPEG~AI \cite{jia2025overview,esenlik2025overview}, FTIC \cite{li2024frequency}, and Cool-Chic \cite{ladune2023cool,ladune2026-coolchic5-0-fasterencoding,balle2025good} have 
emerged as promising alternatives to conventional coding.
These methods use learned components for image representation, coding, and reconstruction, but may differ in their coding frameworks, model architectures, or optimization criteria, including the image quality assessment (IQA) metrics used during training. 
Consequently, learned codecs may produce artifacts that differ from those commonly observed in conventional codecs. They may also alter textures, remove small structures, introduce local inconsistencies, or synthesize plausible but hallucinated details~
\cite{agustsson2019generative,esenlik2025overview,ladune2023cool,jenadeleh2025subjective}.
\begin{figure*}[t!]
\centering
\includegraphics[width=\textwidth,trim=0 4cm 0 5cm,clip]{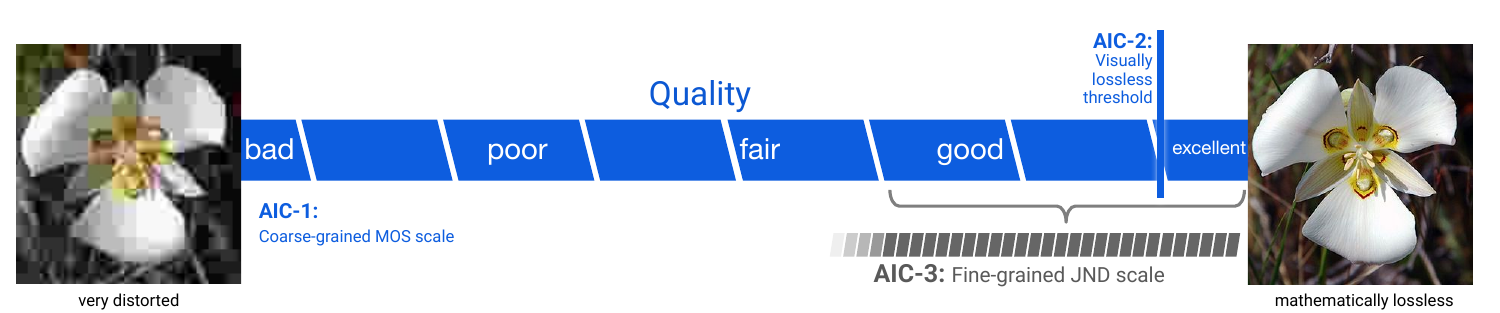}
\vspace{-20pt}
\caption{Schematic overview of the scope of JPEG~AIC-3 and AIC-4.  }
\label{fig_overview_proposed}
\vspace{-10pt}
\end{figure*}

Advances in imaging sensors and display systems have substantially improved spatial resolution and color accuracy. However, storage, bandwidth, and latency constraints make lossy compression essential across a wide range of applications, including digital heritage, remote sensing, e-commerce, professional photography, and medical imaging, where preserving fine image details is critical~\cite{ma2019image}.

High-fidelity lossy compression is therefore required to reduce data volume while preserving fine details relevant to visual assessment and clinical interpretation. At high quality levels, compression artifacts are often subtle, spatially localized, and dependent on image content. They may be masked in textured or complex regions but remain more noticeable in smooth areas, text, faces, edges, and regular structures.

IQA metrics play a central role in the development and evaluation of image codecs, particularly learning-based compression methods. Classical error measures, including the peak signal-to-noise ratio (PSNR) and the mean absolute error (MAE), are simple to compute and have a clear mathematical interpretation that measures the fidelity of the signal, but typically do not correlate well with  perceived image quality. This limits their effectiveness in optimizing image coding for perceptual quality.
Perceptual metrics such as SSIM~\cite{ssim}, MS-SSIM~\cite{ssim}, VMAF~\cite{li2016vmaf}, HDR-VDP-2~\cite{mantiuk2011hdrvdp2}, or CVVDP~\cite{mantiuk2023cvvdp} generally correlate well with subjective quality ratings on datasets such as LIVE~\cite{sheikh2006statistical} and TID2013~\cite{ponomarenko2015tid2013}. These datasets provide mean opinion scores (MOS) for images compressed at coarse distortion levels. However, strong agreement between predicted and subjective MOS in such datasets does not necessarily indicate that a metric can reliably discriminate fine-grained perceptual differences.  This limitation has been observed in previous studies~\cite{zhang2022fine,zhang2019fine, zhang2023perceptual,zhu20242afc,jenadeleh2025subjective,jenadeleh2025fine}, in which metrics that correlate well with subjective scores for coarse distortion levels have been shown to perform poorly when quality differences are subtle.

Hence, developing or benchmarking IQA metrics requires test material covering fine-grained distortion levels, particularly in the high-quality range, and a wide range of compression artifacts. 
The goal is not only to assess strong degradation but also to assess subtle perceptual differences in high-quality images. This requires datasets and evaluation protocols designed for fine-grained quality assessment.

Such datasets should contain high-quality source images with diverse content, including both photographic and non-photographic material, such as natural scenes, portraits, animals, products, screenshots, illustrations, low-light images, and texture-rich regions. This diversity is important because both the visibility of compression artifacts and the performance of IQA metrics depend on image content. The images should also be encoded using a wide range of conventional and learning-based codecs at finely spaced distortion levels to cover the diverse artifacts produced by different codecs and operating points. 
A few studies have already investigated fine-grained IQA \cite{jenadeleh2025subjective,jenadeleh2025fine,zhang2019fine,zhang2022fine,men2021subjective,zhang2023perceptual,richter2009mdct}. However, the available datasets are limited in the number of source images, content diversity, coverage of content types, range of codecs and compression artifacts, quality range, or precision of subjective evaluation.

In response to the growing need to assess subtle perceptual differences, the JPEG committee has launched the AIC-3~\cite{29170-3-AIC3} 
and AIC-4~\cite{jpeg2025aic4} initiatives to establish standards and best practices for the subjective and objective evaluation of full-reference fine-grained IQA. 
Fig.~\ref{fig_overview_proposed} shows the scope of the JPEG~AIC-3 standard~\cite{29170-3-AIC3}. The standard specifies a methodology for fine-grained subjective IQA and the reconstruction of quality scales in fractional JND units, covering the range from good quality to mathematically lossless compression. 
The JPEG~AIC-4 initiative for objective evaluation targets the same quality range. Both AIC-3 and AIC-4 intend to provide fine-grained quality assessment, while previous methodologies such as those described in AIC-1 or BT.500 produce coarse-grained results. In Fig.~\ref{fig_overview_proposed}, the granularity of the scales is reflected in the size of the subdivisions.

In this paper, we introduce AIC2026, a large-scale public dataset for a fine-grained assessment of compressed image quality in the high-quality range, where visual quality differences are subtle, but relevant to modern image coding and applications.

The main contributions of this paper are as follows:
\begin{itemize}

\item We developed a source-image selection procedure based on semantic clustering and inter-metric disagreement analysis to identify diverse and informative source images for image quality assessment. 

\item From 2,787 high-quality candidate images with rich Exif metadata, we selected 70 diverse source images using the proposed procedure followed by manual inspection. These images form the source set of AIC2026 with resolutions ranging from $840{\times}944$ to $3355{\times}2516$.

\item The dataset covers eight conventional and four learning-based image codecs, used in a total of 17 coding configurations, and provides a benchmark for evaluating IQA metrics on artifacts produced by both conventional and emerging learning-based image compression methods.

\item We generated a dense sampling of encodings for each source-codec pair by sweeping relevant codec parameters, and selecting 20 perceptually calibrated distortion levels using CVVDP to cover approximately the range of 0.2 to 4.0 JND in steps of 0.2 JND, supporting a fine-grained quality analysis across codecs with different parameter spaces and distortion-rate behavior.

\item We conducted an extensive objective evaluation of the dataset using 24 conventional and 12 learning-based IQA metrics. The results show that fine-grained quality differences remain challenging for state-of-the-art IQA methods and lead to substantial disagreement among their predictions, particularly for artifacts introduced by learning-based codecs. The inter-metric analysis also shows that disagreement increases as the spacing between distortion levels decreases across all source–codec combinations.
\end{itemize}

To the best of our knowledge, AIC2026 is the first large-scale and diverse dataset for fine-grained compressed images to cover a broad range of artifacts from both conventional and learning-based image codecs, substantially extending the scale and coverage of earlier fine-grained datasets. The dataset is designed as a comprehensive test bed for future subjective studies of the perceptual quality of its compressed images and for objective evaluations of IQA methods. 
Fig.~\historef{} illustrates the scale and granularity of the AIC2026 dataset compared to relevant previous IQA datasets. 

\section{Related Work}
\begin{table*}
\caption{Overview of IQA datasets for estimating the JND threshold and distortion magnitude in fractional JND units.}
\vspace{-5pt}
\label{tb:datasets}
\centering
\begin{threeparttable}

\setlength{\tabcolsep}{1.7pt}
\begin{tabular}{L{3.2cm} C{1.7cm} C{1.7cm} C{1.7cm} C{1.7cm} C{1.5cm}  C{1.7cm} C{1.7cm} C{1.8cm}}
\toprule
Dataset & MCL-JCI & JND-Pano & SIAT-JSSI & Shen~et al. & KonFiG  & KonJND-1k & JPEG~AIC-3 & AIC2026 \\
\midrule
Reference & \cite{MCL-JCI} & \cite{liu2018jnd} & \cite{FanCL_JVCIR19} & \cite{shen2020jnd}  &\cite{men2021subjective}  & \cite{lin2022large} &\cite{testolina2023jpeg,jenadeleh2025subjective} &-- \\
Publication year & 2016 & 2019 & 2020 & 2021 & 2021 & 2022 & 2025 & 2026 \\
\midrule
Image type & regular & panoramic & stereoscopic & regular & regular & regular & regular & regular \\
Number of source images & 50 & 40 & 12 & 202 & 10 & 1,008 & 5 & 70 \\
\midrule
Resolution  of sources & $1920\!\times\!1080$ & $5000\!\times\!2500$ & $1920\!\times\!1080$ & $1920\!\times\!1080$ & $512 \!\times\! 384$ & $640\!\times\!480$ & 
$853\!\times\!945$ to $2592\!\times\!1946$ & $840\!\times\!944$ to $3355\!\times\!2516$ \\
\midrule
Distortion type(s) & JPEG & JPEG & JPEG 2000, HEVC & VVC & JPEG 2000 + 6 artificial& JPEG, BPG & 6 codecs $^{\ast}$& 17 codecs$^{\dagger}$ \\
\midrule
Number of distorted stimuli & 5,000 & 4,000 & 3,498 & 7,878 & 1,020 & 76,104 & 295 & 9,618 \\
\midrule
JND estimation type & threshold & threshold & threshold & threshold & scale & threshold & scale & --- \\ 
\bottomrule
\end{tabular}

\begin{tablenotes}
\footnotesize
\item[$^{\ast}$]  AVIF, JPEG, JPEG~2000, JPEG~XL, VVC Intra, and JPEG~AI.
\item[$^{\dagger}$] 
The above plus HEVC Intra, WebP, PNG (pngquant), Cool-Chic Wasserstein, Cool-Chic MSE, FTIC,  in multiple configurations resulting in 17 codecs.
\\
\end{tablenotes}
\end{threeparttable}
\end{table*}

This section reviews the  most relevant prior work on fine-grained quality assessment of compressed images. We first discuss JND-based image quality datasets that contain images compressed using image codecs or  video codecs in  intra-frame coding. We then review recent work highlighting the need for image datasets with fine-grained distortion levels for evaluating and benchmarking objective IQA metrics. We also discuss related activities within JPEG, which further motivates the development of large-scale datasets of compressed images with diverse content, a broad coverage of conventional and learning-based image codecs and fine-grained distortion levels.

\subsection{JND-based image quality datasets}
\label{sec:jnd_datasets}
JND-based IQA datasets have been widely used to estimate visibility thresholds for visual quality differences between a reference image and its distorted versions. 
In image quality evaluation, one JND is commonly defined as the lowest distortion level at which observers have a 50\%  probability of perceiving the distortion.

Given JND samples from a group of observers, the satisfied user ratio (SUR) at a given distortion level is typically estimated by fitting a normal or Weibull distribution to the estimated JND thresholds and computing the fraction of observers who do not perceive the distortion~\cite{liu2018jnd,shen2020jnd,lin2022large}. 

Several datasets have been developed for JND estimation in image compression. Jin et al.~\cite{MCL-JCI} introduced MCL-JCI, which contains 50 source images at $1920{\times}1080$ resolution and 100 JPEG-compressed versions per source image. JND samples were collected using binary search. Liu et al.~\cite{liu2018jnd} proposed JND-Pano for JPEG-compressed panoramic images viewed with a head-mounted display. Fan et al.~\cite{FanCL_JVCIR19} studied JNDs for stereoscopic images compressed with JPEG~2000 and HEVC Intra coding. Shen et al.~\cite{shen2020jnd} 
 constructed a JND-based image dataset from 202 source images compressed with VVC Intra coding at 39 distortion levels. KonJND-1k increased the scale of JND data collection by using 1,008 source images compressed with JPEG and BPG; JND samples were collected through crowdsourcing using a slider-based adjustment method with a flicker test~\cite{lin2022large}.

These datasets have made important contributions to JND modeling. However, they were used to estimate only the JND threshold. Computing perceived quality above or below the JND threshold from an SUR curve, fitted to JND threshold data collected from a group of subjects, cannot provide  reliable estimates \cite{jenadeleh2024crowdsourced}. 

Distorted versions separated by fractions of a JND are needed to analyze subtle quality differences between consecutive operating points of a source-codec pair. This fine-grained discrimination is important in several use cases, for instance, when objective IQA metrics are used to guide codec optimization, serve as loss functions in learning-based image compression, or evaluate tools such as image enhancement and super-resolution.

\subsection{Fine-grained image quality assessment}
\label{sec:fine_grained_iqa}

Fine-grained IQA aims to measure small perceptual differences between images, rather than only determining whether a distortion is visible. This problem is particularly relevant for high-quality image compression, where the differences between encoded images may be subtle and localized. 

Some studies have investigated fine-grained IQA~\cite{zhang2019fine,zhang2022fine,men2021subjective,zhang2023perceptual,jenadeleh2025subjective,jenadeleh2025fine}; however, they are limited in terms of source image diversity, codec (compression artifact) types, quality range, or the precision of the subjective methodology. 

In~\cite{zhang2019fine}, the authors constructed a dataset of 100 reference images, each compressed at three fixed bitrates (low, medium, and high). For each bitrate, four JPEG quantization strategies were applied by using the typical JPEG table, a uniform table, a PSNR-optimized table, and an MS-SSIM-optimized table, resulting in 1,200 distorted images. To derive perceptual rankings among these subtly distorted images, 30 subjects performed 54,000  pairwise comparisons (PCs). Although the bitrate levels reflect coarse distortions, the differences introduced by the quantization strategies are more subtle. 

In~\cite{zhang2023perceptual}, the authors introduced a dataset of 50 reference images, each compressed at a fixed bitrate of 1.6 bits per pixel (bpp) using 15 perturbed JPEG quantization tables. Over 45,000 PCs were collected to reconstruct quality scores. However, the dataset is restricted to a single codec and bitrate, limiting its applicability to broader compression settings. 

Men et al.~\cite{men2021subjective} presented a dataset of 10 source images, each distorted using seven distortion types: color diffusion, jitter, high sharpening, JPEG~2000 compression, lens blur, motion blur, and multiplicative noise. Each distortion type was applied at 12 levels, yielding 840 distorted images. To increase sensitivity to subtle differences and improve the precision of quality estimation, three artifact boosting techniques were applied: artifact amplification, zooming, and flicker testing. However, only one image compression artifact was used.

Recently, the JPEG~AIC-3 dataset \cite{testolina2023jpeg} was introduced to support the evaluation of image compression methods. The dataset contains five high-quality source images. Each image was first compressed using five lossy image codecs, namely AVIF, JPEG, JPEG~2000, JPEG~XL, and VVC Intra. Later, JPEG AI coded images were also included \cite{jenadeleh2025subjective}. Each source image was encoded at ten operating points. For the subjective study, images were cropped to $620 \times 800$ pixels for side by side comparisons.  The subjective evaluation followed the JPEG~AIC-3 subjective test methodology described in \cite{29170-3-AIC3, jenadeleh2025subjective, jenadeleh2025fine}. Several image quality metrics have also been evaluated and benchmarked on these datasets \cite{mohammadi2026evaluation,jenadeleh2025subjective}. The results show that the performance of objective metrics depends strongly on the image content and on the type of compression artifact. They also show the need for more comprehensive image quality datasets, with a wider variety of image content and a broader range of compression artifacts, including those introduced by learning-based codecs.

Table~\ref{tb:datasets} summarizes existing JND-based IQA datasets. These datasets are limited in source image diversity, source image quality, the range of coding artifacts, or representation of fine-grained  distortion levels. These limitations are important for the evaluation of modern image compression methods. For example, an objective metric that performs well for JPEG artifacts at one bitrate may not show the same behavior for estimating the artifacts produced by some other codecs such as JPEG~XL, AVIF, VVC Intra, or learning-based image compression methods. Therefore, a fine-grained dataset for codec assessment should include a broad set of source-codec pairs, and should sample distortion levels with subtle differences in a meaningful and comparable way across different codecs. 

\subsection{JPEG~AIC activities }
\label{sec:aic_related}
The JPEG Assessment of Image Coding (AIC) project addresses the need for 
sensitive evaluation methods in the high-quality range. The ISO/IEC 29170-2 standard \cite{29170-2-AIC2} (JPEG~AIC-2) defined several tests for visually lossless image coding, one of which is based on a boosting approach using flicker \cite{hoffman2014new}.
It is effective for determining whether a compressed image can be visually distinguished from the reference under flicker-test conditions. However, it does not provide scale values for the perceived quality.

The ISO/IEC 29170-3 standard \cite{29170-3-AIC3} (JPEG~AIC-3) extended the evaluation scope from visually lossless threshold testing to fine-grained subjective assessment, covering the range from good quality to mathematically lossless. The methodology uses plain triplet comparison (PTC) and boosted triplet comparison (BTC). In BTC, operations such as zooming, flickering, and artifact amplification are used to make subtle distortions more visible \cite{men2021subjective}. The collected responses are used to reconstruct perceptual scale values in JND units using a joint model. 

The fourth part of the JPEG~AIC standard will specify benchmark datasets and objective IQA metrics for fine-grained IQA in the same high-fidelity range. 
For this purpose, fine-grained large-scale compressed image datasets are needed, including diverse image content and a broad range of codec artifacts, including those introduced by emerging learning-based methods. The AIC2026 dataset addresses this need.

\section{Source image set generation}
\label{sec_sourceimages}

\subsection{Collecting candidate source images}
To create a diverse set of 70 source images, we first built a large pool of high-quality pristine candidate images. These  were  obtained from photographs captured by volunteer researchers, together with high-quality images from online repositories that allow redistribution and processing for research.

\subsubsection{Images from online public repositories}
The images were obtained from multiple publicly available platforms following the technical  quality and licensing requirements summarized in Table~\ref{tab:criteria}.
\begin{table}
\centering
\caption{Technical and licensing criteria for online sources.}
\vspace{-5pt}
\label{tab:criteria}
\begin{tabular}{@{}p{0.3\linewidth}p{0.65\linewidth}@{}}
\toprule
\textbf{Criterion} & \textbf{Requirement} \\
\midrule
Minimum Resolution & 2000\,$\times$\,3000 pixels \\
Capture Date & On or after 2020-01-01 \\
File Format & Raw format or JPEG (Q{=}96 or higher quality) \\
Metadata & Full EXIF data including camera model, exposure, aperture, ISO, focal length, GPS, and timestamps \\
\bottomrule
\end{tabular}
\end{table}
All downloaded images met these technical requirements. When hosting platforms provided incomplete metadata, all available fields were extracted either from the EXIF metadata or, if accessible, through the platform's API. 
The original download URLs of all retained images are included in the dataset to ensure reproducibility and traceability and to preserve information on licensing terms and source attribution.

\begin{table}[t]
\centering
\caption{Overview of image sources used to construct the candidate image pool.}
\vspace{-5pt}
\label{tab:collected_images}

\adjustbox{width=\columnwidth}{
\setlength{\tabcolsep}{2pt}
\renewcommand{\arraystretch}{1.12}
\small
\begin{tabular}{
p{0.21\textwidth}
R{0.04\textwidth}
p{0.10\textwidth}
R{0.08\textwidth}
R{0.07\textwidth}}
\toprule
\multicolumn{2}{p{0.27\textwidth}}{Origin \hfill Collected} &
Format &
Used in clustering &
Selected \\
\midrule
Wikimedia Commons & 1{,}445 & JPEG \tiny ($Q=100$) & $\checkmark$ & 23 \\
SignatureEdits & 36 & RAW & $\checkmark$ & 4 \\
Photographed natural scenes & 1{,}162 & RAW & $\checkmark$ & 27 \\
Picture of the Year 2024 (Wikimedia) & 88 & JPEG \tiny ($Q\geq96$) & -- & 9 \\
JPEG~AI synthetic set & 36 & PNG & -- & 2 \\
Photographed products & 18 & RAW & -- & 3 \\
Screenshots & 2 & PNG & -- & 2 \\
\midrule
\textbf{Total} & \textbf{2{,}787} & &  \textbf{2{,}643}& \textbf{70} \\
\bottomrule
\end{tabular}
}
\vspace{-10pt}
\end{table}

\subsubsection{Images photographed by contributors}
An additional set of 1{,}180 images was provided by researchers and collaborators under the CC0 license, including 1{,}162 natural scenes and 18 product images. All images were captured at native camera resolution and stored in both RAW format and the highest quality JPEG format supported by each camera. The high-quality JPEG versions were used for candidate-image clustering and source selection, whereas the corresponding RAW files were processed to produce the final source images.

The images cover a wide range of subjects, including individuals, objects, natural environments, and indoor settings. 
Data collection was conducted in compliance with applicable EU data protection regulations (e.g., GDPR). Consent was obtained from individuals photographed. 

Table~\ref{tab:collected_images} summarizes the number and format of the images collected and selected for the AIC2026 dataset.

\subsection{Semantic clustering of candidate images}
To support the selection of a diverse source set, we extracted deep visual features from the candidate images and clustered them using K-means. 
The candidate images were grouped into 60 clusters. This allowed multiple images to be selected from clusters containing content of particular interest, such as portraits and animals, while leaving room for the manual inclusion of underrepresented content types, including product images, illustrations, and screen content.

\begin{figure}[t!]
\centering
\includegraphics[width=1\columnwidth]{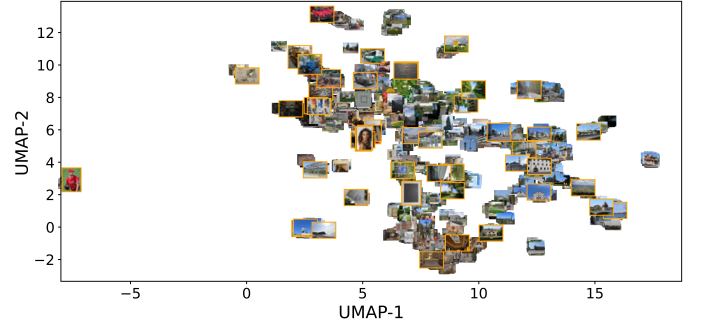}
\vspace{-15pt}
\caption{
2D UMAP projection of deep visual embeddings for candidate images. Selected images are highlighted. 
}
\label{fig:umap_all}
\vspace{-10pt}
\end{figure}

The pretrained DINOv2 ViT-L/14 vision transformer~\cite{oquab2024dinov2} was used as a frozen feature extractor because 
it provides a visual  representation that captures image similarity without relying on predefined categories. 
For the purpose of clustering only, each image was resized so that its shorter side was 518 pixels, center-cropped to $518 \times 518$ pixels, converted to RGB, and normalized using the ImageNet mean and standard deviation. The class-token embedding was used as a 1{,}024-dimensional image-level descriptor. The resulting descriptors for the 2{,}643 candidate images were then grouped into 60 clusters using K-means to guide diversity-aware source selection. 
 Fig.~\ref{fig:umap_all} shows the two-dimensional UMAP projection of deep visual embeddings for the candidate images.  

The candidate nearest the cluster centroid is not necessarily the most informative source image, particularly in clusters dominated by repetitive content such as sky, sea, foliage, or other largely uniform textures. To better capture within-cluster variability, we selected 10 representative images at approximately equal Euclidean-distance intervals from the centroid, resulting in a pool of 600 candidate images that spans both central and peripheral regions of the feature space. For the 30 clusters with  odd identifiers, \(i = 1, 3, 5, \ldots, 59\), the candidates were ordered by their cosine distance from their corresponding  cluster centroid, and for the $i$-th cluster the image at distance rank $1 + \lfloor i/2 \rfloor \bmod 10$ was selected. For the remaining 30 clusters, the representative image was selected using the inter-metric disagreement analysis described in the next subsection. Additional images were retained from some clusters containing content of particular interest, such as portraits and animals.

\vspace{-10pt}
\subsection{Inter-metric disagreement analysis}
\label{metrics_disagreement}
To increase dataset diversity with respect to codec-induced distortions and to identify images that are informative for evaluating objective image quality metrics, we analyzed inter-metric disagreement (IMD) within the remaining 30 clusters, with even identifiers.

{\em Definition.}
Let $M=\{1,\ldots,n\}$ index the set of $n$ objective quality
metrics. For each source image $s$, let $\mathbf{m}_i^s$ denote the
vector of quality scores assigned by metric $i\in M$ to the
corresponding distorted images. The inter-metric disagreement score
$\imd_{M}(s)\in[0,1]$ is defined as
\begin{equation}
\imd_{M}(s)
= \mkern-3mu
\frac{1}{2\binom{n}{2}}\mkern-7mu
\sum_{1\leq i<j\leq n}\mkern-12mu
\bigl(
2-
\lvert\rho(\mathbf{m}_i^s,\mathbf{m}_j^s)\rvert-
\lvert\tau(\mathbf{m}_i^s,\mathbf{m}_j^s)\rvert
\bigr)
,
\label{eqn:metric_disagreement}
\end{equation}
where $\rho$ and $\tau$ denote Spearman's rank-order correlation
coefficient (SRCC) and Kendall's rank correlation coefficient (KRCC), respectively.

We quantified metric disagreement using only rank-based measures, using absolute values to account for metrics with differing polarity, i.e. scores reflecting quality (higher is better) or distortion (lower is better). 
A higher value of \(\imd\limits_M(s)\) indicates stronger disagreement between the  metrics, identifying images for which subjective scores are expected to be more informative for the metric evaluation.

Each candidate image was compressed using four representative codecs: JPEG, JPEG~2000, JPEG~XL, and AVIF. For each codec, seven encoder-defined quality settings were used, resulting in 28 distorted versions per image.
For each distorted image, seven objective IQA metrics were computed:
$M=\{$
PSNR-Y, SSIM~\cite{ssim}, MS-SSIM~\cite{ssim}, VMAF-neg~\cite{vmaf-neg}, Butteraugli-pnorm~\cite{butteraugli}, CVVDP~\cite{mantiuk2023cvvdp}, and SSIMULACRA2~ \cite{sneyers2022ssimulacra2} $\}$. These are the anchor metrics listed in the common test condition (CTC) document for AIC-4 \cite{aic_ctc}. 
This limited selection of codecs, encoder settings, and metrics reduces the computational cost of calculating IMD for all candidate images, while still acting as a proxy for the IMD computed over more codecs, distortion levels, and metrics.
For clusters with even identifiers, \(0, 2, 4, \ldots, 58\), the candidate with the largest inter-metric disagreement score was selected.

\subsection{Inclusion of additional content types}
All images selected using the procedure described above were manually inspected. Images with visible quality defects were replaced with alternatives of similar content, whereas images containing potentially sensitive or inappropriate material were replaced with suitable candidates from the image pool. When a cluster was already adequately represented, the replacement was drawn from a cluster requiring broader coverage, such as the portrait cluster. More samples were added from the portrait and animal clusters. In total, 63 images were selected.

Because most images obtained from Wikimedia Commons or contributed by researchers depicted natural scenes, an additional set of 58 candidate images was collected to broaden the range of content types. Following manual inspection, seven source images were added: two illustrations, three product images, and two screenshots.

The final dataset comprises 70 pristine source images selected to provide broad content diversity, include images with high inter-metric disagreement, and ensure coverage of relevant scene and content types.

\begin{table}[t]
\centering
\caption{Source image categories.}\label{tab:categories}
\vspace{-5pt}
\begin{tabular}{l|c|c|c}
\toprule
Category & Resampled & Native (1:1) & Total\\
\midrule
1 (cropped)    & 5	& 5  & 10\\
2 (downscaled) & 28	& 0  & 28\\
3 (larger images)   & 22	& 10 & 32\\
\bottomrule
\end{tabular}
\vspace{-10pt}
\end{table}
\begin{figure*}[ht!]
\centering
\includegraphics[width=0.98\textwidth]{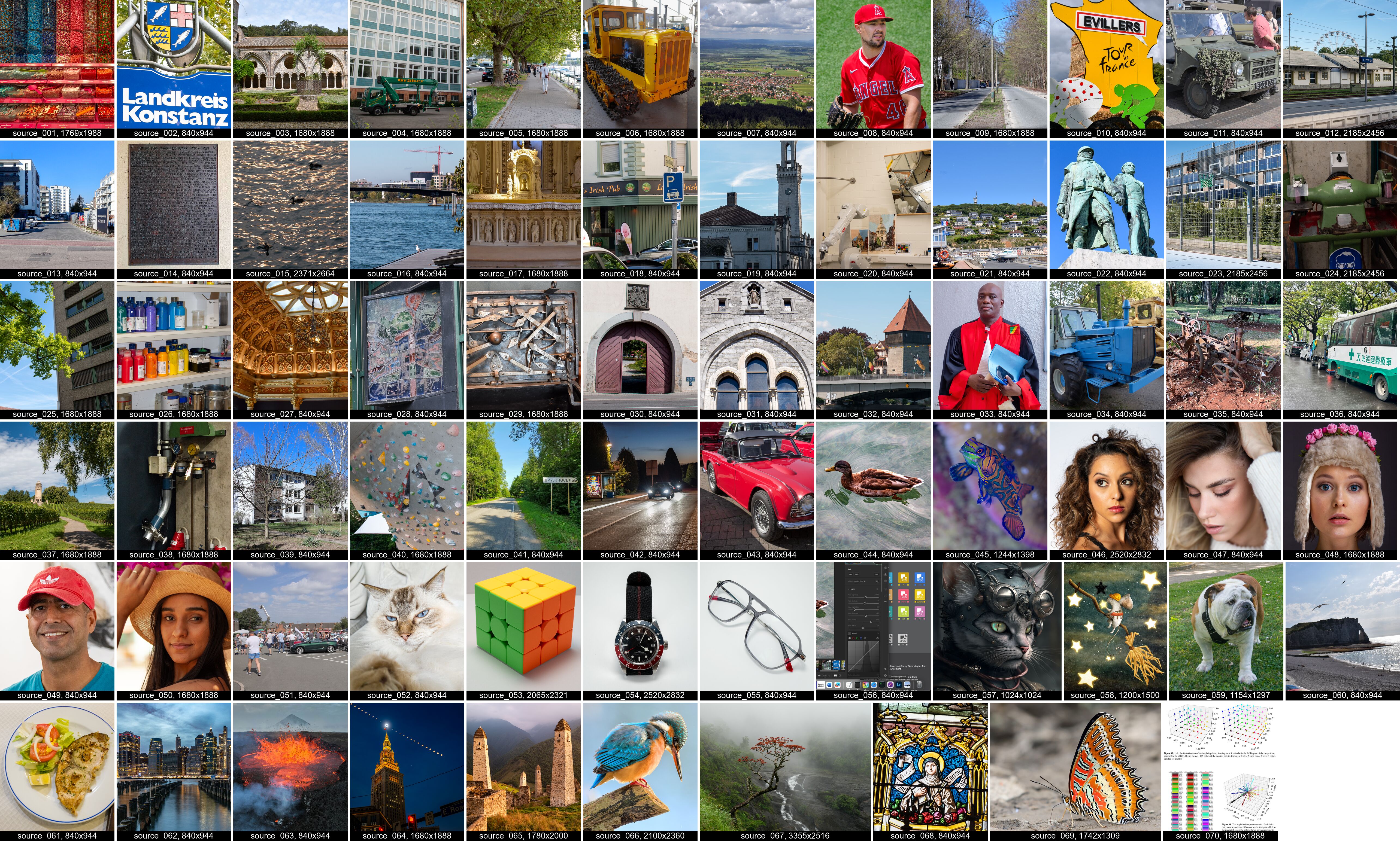}
\caption{Thumbnails of the 70 source images.}
\label{fig:thumbnails}
\vspace{-10pt}
\end{figure*}
\subsection{Processing of selected source images}
The 70 selected candidates were manually processed with Adobe Lightroom. 
Two JPEG sources S35 and S36, which are in the Display~P3 color space, were  converted to sRGB; the other JPEG sources were already in sRGB. 
In the case of RAW sources, exposure, contrast, color and tone adjustments were applied, primarily using the auto-adjustment suggested by Lightroom; for some images, additional global corrections were applied, such as dehazing.

Three categories were defined for cropping and downscaling,
corresponding to common image processing workflows:
\begin{enumerate}
    \item The original candidate image was cropped substantially in both dimensions to obtain a \(840 \times 944\) source image.
   \item The original candidate image was downscaled and then cropped along only one dimension to adjust the aspect ratio and obtain a source image of \(840 \times 944\) pixels.
    \item The source image is larger than \(840 \times 944\) pixels, and has dimensions up to $3355{\times}2516$ pixels.
\end{enumerate}

For the 32 original images that were available only in JPEG format, 
a downscaling factor of at least 2 was applied to mitigate possible compression artifacts, even though only high-quality JPEG images ($Q \in [96,100]$) were selected.
From the 38 other original images that were available in RAW or PNG format,
15 images were retained in their native capturing resolution~(1:1); the others were also downscaled.

The processed images were stored in the sRGB color space in 8-bit precision, using lossless PNG files,
ensuring that all codecs can handle the input format correctly and that in browsers (e.g., for a crowdsourced subjective experiment), all images will be displayed correctly and without gamut clipping.

Table~\ref{tab:categories} gives a breakdown of the three categories.
Fig.~\ref{fig:thumbnails} presents thumbnails of the final set of 70 source images. 

\section{Distorted image set generation}
This section describes the encoding settings and procedures used to generate the compressed versions of the source images. Each source image was encoded using seven codecs or coding configurations at 20 approximately perceptually uniform distortion levels. Overall, the dataset includes 17 codecs or coding configurations derived from eight conventional and four learning-based codecs, providing broad coverage of compression artifacts across the dataset.

\subsection{Encoding and estimation of distortion levels}
\label{Encoding_and_estimation_distortion_levels}
Each source image was encoded with the corresponding codec or codec configuration by sweeping the relevant encoder settings from the highest-quality operating point to settings that produced lower bitrates.
Depending on the codec, these settings included quality factors, quantization parameter (QP), target bitrates, the distortion-rate trade-off parameter ($\beta$) used in learning-based codecs, coding modes, or other encoder options that affect the distortion-rate behavior. The sweep was performed densely to obtain candidate images with small quality differences between successive compressed versions.  The compressed images were then evaluated using the CVVDP metric \cite{mantiuk2023cvvdp}, and the resulting scores were mapped to JND units.  CVVDP was  selected because it showed the highest overall correlation with subjective scores in the AIC-3 dataset~\cite{jenadeleh2025subjective}. 
If a different metric were selected, the selected target bitrates would differ slightly. Nevertheless, this selection does not favor CVVDP in the performance evaluation.

CVVDP scores were computed using the standard Full HD SDR display configuration ({\small \verb|cvvdp -d standard_fhd|}). This preset defines a fixed viewing model for SDR IQA, with an effective angular resolution of 37.84 pixels per degree, a peak luminance of 200 \(\mathrm{cd}/\mathrm{m}^{2}\), a black level of 0.2 \(\mathrm{cd}/\mathrm{m}^{2}\), and a reflected-light level of 0.3979 \(\mathrm{cd}/\mathrm{m}^{2}\). 
The resulting CVVDP scores, expressed on a 0-10 quality scale with 10 indicating the highest quality, were converted to JND units using a mapping estimated from the subjective scores of the AIC-3 dataset:
\begingroup
\setlength{\abovedisplayskip}{3pt}
\setlength{\belowdisplayskip}{3pt}
\begin{equation}\label{CVVDPtoJND}
\textrm{CVVDP}_{\textrm{JND}} = 3.1889(10-\textrm{CVVDP})^{1.0129}.
\end{equation}
\endgroup
The parameters were estimated by nonlinear least-squares fitting. 
This mapping reflects not only the conversion from a quality scale to a distortion scale (for which simply $10-x$ would suffice), but also the more sensitive viewing conditions and test protocols used in the AIC-3 methodology, compared to the JOD units that  CVVDP was calibrated to.

Finally, for each source--codec pair, 20 compressed images were selected to cover the target distortion range of 0.2--4.0 JND at intervals of approximately 0.2 JND. 
The upper limit of 4.0 JND was selected to fully cover the approximate 0--3 JND range targeted by the JPEG AIC-3 standard~\cite{29170-3-AIC3}, while allowing additional headroom for possible CVVDP prediction errors, particularly for compression artifacts that differ from those represented in its calibration data. Because the AIC-3 subjective data span approximately within 0--2.5 JND, mapped values beyond this range are extrapolated.  
Also, Hammou et al. \cite{hammou2026evaluating} showed that CVVDP does not fully capture contrast constancy for supra-threshold distortions, limiting its accuracy for higher distortion levels.

\begin{figure*}[ht!]
\centering
\includegraphics[width=0.95\textwidth]{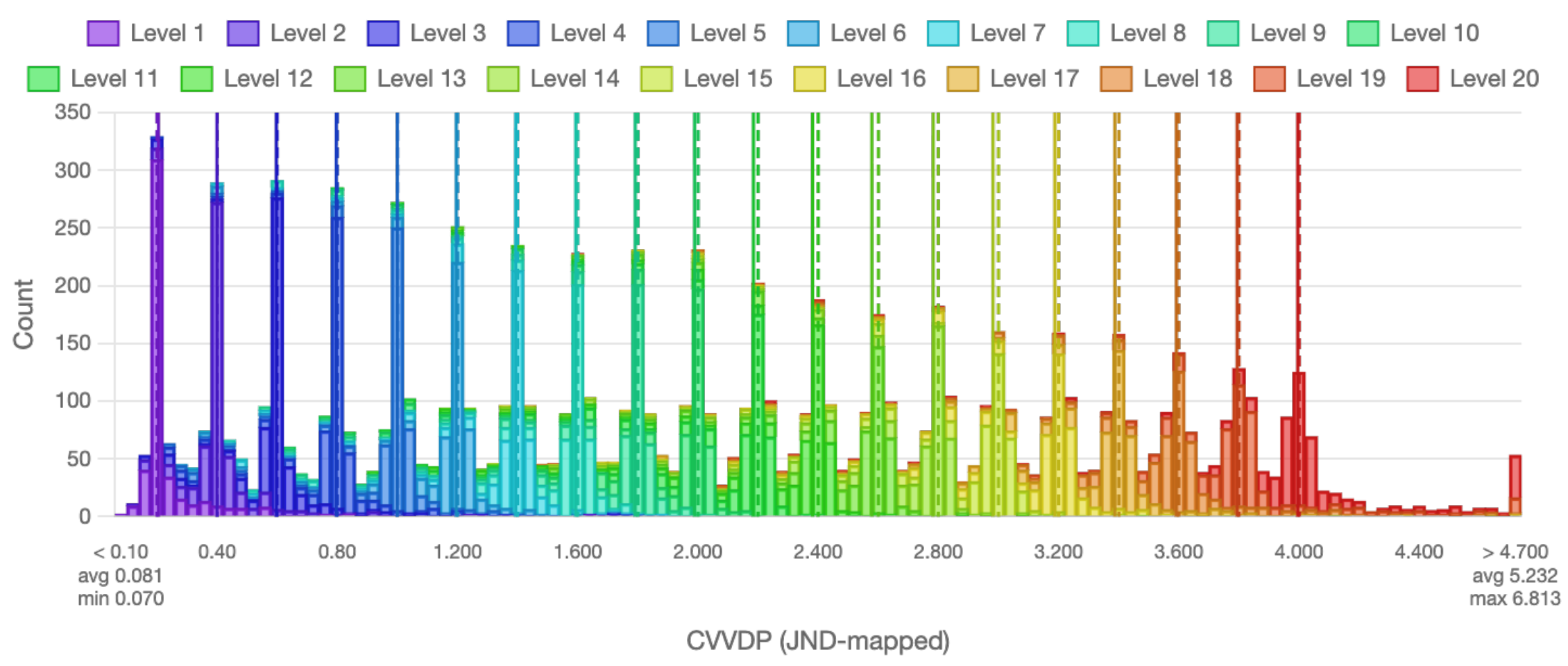}
\vspace{-8pt}
\caption{Histogram of JND-mapped CVVDP scores of the distorted images in the AIC2026 dataset, colored by distortion level.
Dashed vertical lines indicate the target score for each distortion level; solid lines indicate the median achieved value.}
\label{fig:histo_cvvdp}
\end{figure*}

When fewer than 20 distinct compressed images were available within this range, additional candidates outside the range were included. These supplementary samples were selected by balancing candidates below 0.2 JND and above 4.0 JND to maintain coverage near the target quality range. 
This situation occurred more frequently using video codecs, whose encoder settings are often limited to a finite set of QPs, e.g., 51 or 63 values. 
For some source-codec pairs, even the highest-quality encoder settings did not yield compressed images at the lowest target distortion levels, such as 0.2 JND. This was observed mainly for learning-based codecs, WebP, and color-reduced PNG. In the case of WebP, the limitation was partly related to the use of 4:2:0 chroma subsampling, while for color-reduced PNG, a 256-color palette was not always sufficient to represent the source image at very low distortion. Despite these codec-specific constraints, the selected compressed images closely followed the target CVVDP-estimated JND levels across the dataset, as shown in Fig.~\ref{fig:histo_cvvdp}.

Although codec comparison is not the objective of this work, experts from several codec development teams were consulted to review the encoding settings.
To support reproducibility, we have included the complete encoding recipes for all codecs in the dataset repository.
The following coding configurations were used:
\\
Base codec set:
\begin{enumerate}
\item \textbf{JPEG}: libjpeg-turbo 2.1.4-2 ImageMagick~6.9.12 defaults.
\item \textbf{JPEG~2000}: Kakadu 8.4.1 with perceptual tuning.
\item \textbf{JPEG~XL}: libjxl 0.11.1, cjxl defaults.
\item \textbf{AVIF}: aom 3.12.1, avifenc 1.3.0 defaults.
\item \textbf{JPEG~AI}, high operation point (HOP). %
See Section~\ref{learning_based_codecs}.
\end{enumerate}
Extended codec set:
\begin{enumerate}
\setcounter{enumi}{5}
\item \textbf{JPEG}: \textbf{reference} libjpeg 1.71, `on steroids' options~\cite{richter2016jpeg}.
\item \textbf{JPEG}: \textbf{Jpegli} 0.11.1 (perceptually optimized encoder).
\item \textbf{JPEG~XL}:  \textbf{Modular} mode instead of VarDCT.
\item \textbf{AVIF}: avifenc with {\small \bf \tt tune=iq}.
\item \textbf{WebP}: libwebp 1.2.4, cwebp defaults. 
\item \textbf{PNG}: pngquant 2.14.1 (color reduction) and oxipng 9.1.5. 
\item \textbf{JPEG~AI}, base operation point (BOP). 
See Section~\ref{learning_based_codecs}.

\item \textbf{HEVC}: HM 16.20 with Screen Content Coding (SCM~8.8) extensions enabled, as specified in the JPEG~AI common training and test conditions (CTTC) v8.0~\cite{jpegai_cttc_v8_2023}.

\item \textbf{VVC}: VTM with Screen Content Coding (SCC) enabled, as specified in 
\cite{jpegai_cttc_v8_2023}.

\item \textbf{Cool-Chic} 4.2, optimized for Wasserstein distortion~\cite{balle2025good}.
\item \textbf{Cool-Chic} 4.1, optimized for MSE \cite{ladune2023cool}.
\item \textbf{FTIC}, optimized for MS-SSIM \cite{li2024frequency}.
\end{enumerate}
The encoding recipes for the learning-based image compression methods, corresponding to codecs~5, 12, and 15--17, are detailed in the following subsection.
\subsection{Learning-based image compression methods}
 \phantomsection
\label{learning_based_codecs}
 We considered only learning-based image compression methods capable of producing a complete and decodable bitstream, including all required syntax elements and side information. We selected four methods that satisfy these requirements: JPEG~AI, Cool-Chic optimized for MSE, Cool-Chic optimized for Wasserstein distortion, and FTIC. 
 
\begin{enumerate}
\setcounter{enumi}{4}
\item \textbf{JPEG~AI (HOP):} Compressed images were generated using the JPEG~AI reference software \cite{jpegai-referencesoftware} with \texttt{high} profile and by disabling the automatic bitrate matcher and explicitly specifying the model index and beta-displacement log value. 
Each source image was encoded using all combinations of the four models and 37 operating points obtained by sampling the beta-displacement parameter over its full range, including the minimum and maximum values, resulting in $4 \times 37 = 148$ compressed images per source.
\end{enumerate}

\begin{enumerate}
\setcounter{enumi}{11}

\item \textbf{JPEG~AI (BOP)}: 
The \texttt{base} profile and all tools were disabled. 
The same procedure as that used for the HOP configuration was then applied, resulting in 148 compressed images generated for each source image. 
\end{enumerate}
For the HOP and the BOP configurations of JPEG AI, 20  compressed versions per source image were selected in two stages. In the first stage, we identified the global distortion–rate Pareto front and retained only points not dominated by any other operating point.  In the second stage, 20 operating points, corresponding to 20 compressed images, from this subset were selected to span the range from 0.2 to 4.0 JND at steps of approximately 0.2 JND, as described in Section~\ref{Encoding_and_estimation_distortion_levels}.

\begin{figure}[!t]
\begin{minipage}[t]{0.24\textwidth}
\centering
\includegraphics[width=\linewidth]{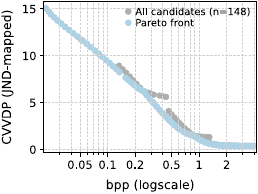}
\centerline{(a)}
\end{minipage}
\hfill
\begin{minipage}[t]{0.24\textwidth}
\centering
\includegraphics[width=\linewidth]{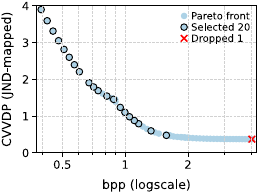}
\centerline{(b)}
\end{minipage}
\vspace{-5pt}
\caption{Selected distorted images and distortion-rate results for source~03 using JPEG AI (BOP). (a) All 148 candidates, with Pareto-front samples shown in blue. (b) Pareto-front samples only; selected samples are circled in black. The highest-bitrate sample, marked by a red~$\times$, was excluded because it gives only a very small perceptual gain at a very high bitrate.}
\label{fig:source003_results}
\end{figure}

The perceptual quality gain appears to saturate as the bitrate increases. Therefore, the sample with the highest bitrate was excluded.
Fig.~\ref{fig:source003_results} exemplifies this procedure. 

\begin{enumerate}
\setcounter{enumi}{14}
\item 
\textbf{Cool-Chic 4.2 (Wasserstein)}~\cite{balle2025good,cool-chic-software} is an overfitted image compression model in which the latent representations, the synthesis network, and the entropy network are optimized jointly for each source image by minimizing a distortion-rate objective of the form \(R + \lambda D\).   The latent representations are stored as multi-resolution spatial arrays that are progressively upsampled and concatenated during decoding, and a lightweight convolutional synthesis network maps them to the reconstructed image. The entropy network models the conditional distribution of latent elements to enable entropy coding. The resulting bitstream therefore consists of the entropy-coded latent arrays together with the quantized parameters of the synthesis and entropy networks, reflecting the per-image optimization of the entire codec. Cool-Chic 4.2  optimizes for Wasserstein Distortion.

A total of 41 compressed images per source image were generated using logarithmically spaced values of $\lambda$ ranging from $10^{-6}$ to $10^{-2}$. For all encodings,  the \texttt{perceptive} encoder configuration, the \texttt{hop} residual decoder configuration, and the Wasserstein tuning objective, were used. The $\lambda$ value was varied across the specified range, while all other settings were kept fixed. A total of 20 of the 41 compressed images per source were selected. Whenever possible, the selected images covered the range from 0.2 to 4.0 JNDs at approximately 0.2-JND intervals.

\item \textbf{Cool-Chic 4.1 (MSE)}: 
For the Cool-Chic MSE codec~\cite{ladune2023cool,cool-chic-software}, each source image was encoded using the \texttt{medium_30k} encoder configuration while sweeping $\lambda$. The sweep produced 41 compressed images per source, from which again 20 were selected as for the Cool-Chic Wasserstein codec. 

\end{enumerate}

\begin{figure}
\centering
\includegraphics[width=0.65\linewidth]{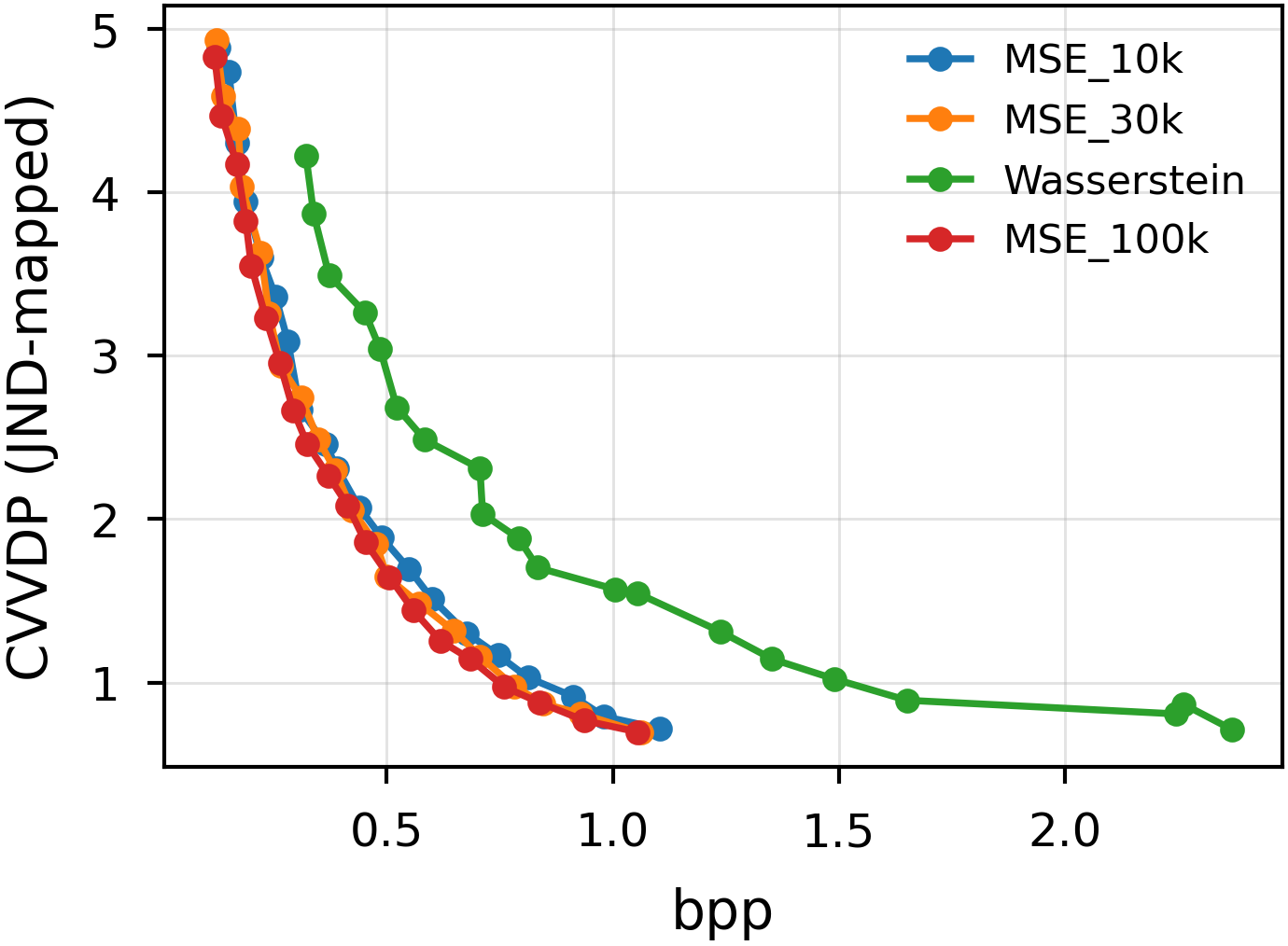}
\vspace{-5pt}
\caption{ Distortion-rate curves for source 10 obtained with Cool-Chic MSE and Cool-Chic Wasserstein objectives.}
\label{fig:CCM10}
\end{figure}

For source images in the AIC2026 dataset, Cool-Chic optimized for MSE typically yields smoother distortion–rate curves when evaluated with conventional image quality metrics such as PSNR, MS-SSIM, or CVVDP. 
Fig.~\ref{fig:CCM10} illustrates this trend for source image~10 by showing the estimated distortion-rate curves obtained using different parameter settings for Cool-Chic (MSE) and Cool-Chic (Wasserstein). 
The differences between the two compression methods become particularly evident at low bitrates. Cool-Chic (MSE) tends to produce blurred or locally averaged regions, whereas Cool-Chic (Wasserstein) may produce repeated patterns and small spatial displacements of fine details. According to the Cool-Chic (Wasserstein) authors, the repeated patterns may be related to the autoregressive model used to code the latent representation, while the spatial displacements are consistent with the design of the Wasserstein distortion, which is intentionally insensitive to small translations and texture resampling.

Because Cool-Chic (MSE) and Cool-Chic (Wasserstein) produce distinct types of compression artifacts, both methods were retained in the proposed dataset to increase artifact diversity and to better represent the range of distortions generated by modern learned image compression methods.

\begin{enumerate}
\setcounter{enumi}{16}
\item \textbf{FTIC (MS-SSIM)}  is a frequency-aware transformer-based image compression method proposed in~\cite{li2024frequency}. It introduces a frequency-aware transformer block that uses a frequency-decomposition window attention to capture low-frequency, high-frequency, horizontal, and vertical image components. These components are further modulated by a frequency-modulation feed-forward network operating in the frequency domain. In addition, FTIC employs a transformer-based channel-wise autoregressive entropy model to exploit inter-channel dependencies in the latent representation and improve entropy estimation.

Due to the high computational cost of training, we used the six pretrained FTIC models released by the authors~\cite{ftic-software}.
In our experiments, only the MS-SSIM optimized models were used. Each source image was encoded once with each pretrained model, producing six compressed images per source image.%

\end{enumerate}

\subsection{Codec assignment}

Each source image is encoded with all five base codecs and two codecs from the extended set.
Table~\ref{tbl:codec_usage} lists the source images assigned to each extended codec. This source-codec assignment was designed to maximize diversity of source-codec combinations. It provides broad codec coverage and supports analysis of codec-dependent quality degradation and artifact characteristics across diverse image content.

\begin{table}[tb]
\caption{Codec assignments showing all  codecs and respective configurations. Base codecs (1--5) are used for all 70 sources, while extended codecs (6--17) are assigned to subsets.
}
\label{tbl:codec_usage}

\setlength{\tabcolsep}{2pt}
\begin{tabular}{rllc|l}
\toprule
id & acronym & codec & \# & Source indices \\
\midrule
1 & \verb|JPG| & JPEG~\cite{jpeg} & 70 & all \\
2 & \verb|J2K| & JPEG~2000~\cite{rabbani2002overview}& 70 & all \\
3 & \verb|JXL| & JPEG~XL~\cite{jxl2025}  & 70 & all \\
4 & \verb|AVIF| & AVIF~\cite{barman2020evaluation} & 70 & all \\
5 & \verb|JAI| & JPEG~AI (HOP)~\cite{esenlik2025overview} & 70 & all \\
\midrule
6 & \verb|JPGR| & JPEG-ref~\cite{richter2016jpeg}& 11 & \tiny
07, 08, 24, 26, 29, 30, 38, 42, 44, 53, 59 \\
7 & \verb|JPGL| & Jpegli~\cite{bruse2024users} & 11 & \tiny
11, 12, 13, 16, 22, 26, 50, 58, 62, 65, 68 \\
8 & \verb|JXLM| & JPEG~XL modular~\cite{jxl2025}  & 11 & \tiny
01, 08, 21, 23, 27, 28, 39, 52, 54, 55, 58 \\
9 & \verb|AVIFQ| & AVIF tune-iq~\cite{barman2020evaluation}& 12 & \tiny
05, 11, 15, 17, 36, 37, 39, 42, 43, 46, 57, 63 \\
10 & \verb|WEBP| & WebP~\cite{libwebp124} & 12 & \tiny
03, 09, 14, 17, 19, 20, 25, 29, 36, 45, 54, 65 \\
11 & \verb|PNG| & PNG~\cite{pngquant2141} & 11 & \tiny
04, 13, 15, 20, 23, 31, 32, 33, 38, 41, 51 \\
12 & \verb|JAIB| & JPEG~AI (BOP)~\cite{esenlik2025overview} & 11 & \tiny
03, 04, 05, 28, 40, 47, 53, 61, 64, 67, 68 \\
13 & \verb|HEVC| & HEVC Intra~\cite{sullivan2012overview} & 12 & \tiny
09, 10, 31, 34, 35, 48, 52, 59, 61, 62, 63, 66 \\
14 & \verb|VVC| & VVC Intra~\cite{hamidouche2022versatile} & 12 & \tiny
01, 02, 06, 14, 32, 35, 44, 50, 57, 64, 66, 70 \\
15 & \verb|CCW| &  Cool-Chic Wasserstein~\cite{balle2025good}  & 12 & \tiny
06, 10, 18, 22, 30, 45, 46, 47, 49, 51, 55, 60 \\
16 & \verb|CCM| & Cool-Chic MSE \cite{ladune2023cool}& 12 & \tiny
02, 07, 16, 18, 19, 27, 33, 34, 43, 56, 67, 69 \\
17 & \verb|FTIC| & FTIC MS-SSIM \cite{li2024frequency}& 13 & \tiny
12, 21, 24, 25, 37, 40, 41, 48, 49, 56, 60, 69, 70 \\
\bottomrule
\end{tabular}
\end{table}

\section{Objective metric results and analysis}
\label{sec:objective_metrics}

\begin{table*}
\centering
\caption{Inter-metric correlations between various full-reference IQA metrics, for the complete AIC2026 dataset. Below the diagonal: $|\mathrm{SRCC}|$ and  
above the diagonal: $|\mathrm{KRCC}|$ values.
}
\vspace{-5pt}
\label{tab:metrics}
\adjustbox{width=\textwidth}{
\setlength{\tabcolsep}{2pt}

\begin{tabular}{lllcl||ccccccccccccccccccccccccc||ccccccccccc}
\toprule
& & & & & \multicolumn{23}{c||}{\textit{Conventional metrics}} &
\multicolumn{12}{c}{\textit{Learning-based metrics}} \\
Metric & Year & Description & $\uparrow$/$\downarrow$ & Ref. & \rotatebox{90}{\strut PSNR} & \rotatebox{90}{\strut PSNR-Y} & \rotatebox{90}{\strut PSNR-HVS} & \rotatebox{90}{\strut mDCTPSNR} & \rotatebox{90}{\strut SSIM} & \rotatebox{90}{\strut MS-SSIM} & \rotatebox{90}{\strut SSIMU2} & \rotatebox{90}{\strut DSSIM} & \rotatebox{90}{\strut IW-SSIM} & \rotatebox{90}{\strut FSIM} & \rotatebox{90}{\strut VMAF v0} & \rotatebox{90}{\strut VMAF-neg v0} & \rotatebox{90}{\strut VMAF v1} & \rotatebox{90}{\strut VIF} & \rotatebox{90}{\strut ADM2} & \rotatebox{90}{\strut CIEDE2000} & \rotatebox{90}{\strut VSI} & \rotatebox{90}{\strut GMSD} & \rotatebox{90}{\strut NLPD} & \rotatebox{90}{\strut HaarPSI} & \rotatebox{90}{\strut FLIP} & \rotatebox{90}{\strut HDR-VDP-2} & \rotatebox{90}{\strut HDR-VDP-3} & \rotatebox{90}{\strut CVVDP} & \rotatebox{90}{\strut LPIPS-alex} & \rotatebox{90}{\strut STLPIPS} & \rotatebox{90}{\strut PieAPP} & \rotatebox{90}{\strut WaDIQaM} & \rotatebox{90}{\strut DISTS} & \rotatebox{90}{\strut AHIQ} & \rotatebox{90}{\strut DeepDC} & \rotatebox{90}{\strut DreamSim} & \rotatebox{90}{\strut TOPIQ} & \rotatebox{90}{\strut WD} & \rotatebox{90}{\strut A-FINE} & \rotatebox{90}{\strut DMM} \\
\midrule

PSNR         & & Peak Signal-to-Noise Ratio (RGB)                    & $\uparrow$ & & & \cellcolor[HTML]{a3f679}$.83$ & \cellcolor[HTML]{d9f679}$.62$ & \cellcolor[HTML]{f6eb79}$.46$ & \cellcolor[HTML]{f6d179}$.35$ & \cellcolor[HTML]{f6e779}$.44$ & \cellcolor[HTML]{ebf679}$.54$ & \cellcolor[HTML]{e2f679}$.58$ & \cellcolor[HTML]{f6ce79}$.34$ & \cellcolor[HTML]{f6cc79}$.33$ & \cellcolor[HTML]{f6e479}$.43$ & \cellcolor[HTML]{f6e779}$.44$ & \cellcolor[HTML]{f6e779}$.44$ & \cellcolor[HTML]{f6d879}$.38$ & \cellcolor[HTML]{f6e679}$.43$ & \cellcolor[HTML]{c2f679}$.71$ & \cellcolor[HTML]{f6ef79}$.47$ & \cellcolor[HTML]{f4f679}$.51$ & \cellcolor[HTML]{ccf679}$.67$ & \cellcolor[HTML]{e3f679}$.58$ & \cellcolor[HTML]{d6f679}$.63$ & \cellcolor[HTML]{f6d579}$.37$ & \cellcolor[HTML]{f6d679}$.37$ & \cellcolor[HTML]{f6f079}$.47$ & \cellcolor[HTML]{f6c579}$.30$ & \cellcolor[HTML]{f6d979}$.38$ & \cellcolor[HTML]{f6e079}$.41$ & \cellcolor[HTML]{f69c79}$.14$ & \cellcolor[HTML]{f6ce79}$.34$ & \cellcolor[HTML]{f6ac79}$.20$ & \cellcolor[HTML]{f6bb79}$.26$ & \cellcolor[HTML]{f6d479}$.36$ & \cellcolor[HTML]{f6e179}$.41$ & \cellcolor[HTML]{f6ef79}$.47$ & \cellcolor[HTML]{f6d079}$.35$ & \cellcolor[HTML]{f6f679}$.50$ \\
PSNR-Y       & & PSNR on Y luma channel            & $\uparrow$ & & \cellcolor[HTML]{84f679}$.96$ & & \cellcolor[HTML]{dbf679}$.61$ & \cellcolor[HTML]{f1f679}$.52$ & \cellcolor[HTML]{f6d779}$.37$ & \cellcolor[HTML]{f4f679}$.51$ & \cellcolor[HTML]{e2f679}$.58$ & \cellcolor[HTML]{d6f679}$.63$ & \cellcolor[HTML]{f6de79}$.40$ & \cellcolor[HTML]{f6d079}$.35$ & \cellcolor[HTML]{f6e979}$.45$ & \cellcolor[HTML]{f6ec79}$.46$ & \cellcolor[HTML]{f6e979}$.45$ & \cellcolor[HTML]{f6e679}$.44$ & \cellcolor[HTML]{f6ed79}$.46$ & \cellcolor[HTML]{d4f679}$.64$ & \cellcolor[HTML]{f6e879}$.44$ & \cellcolor[HTML]{ebf679}$.55$ & \cellcolor[HTML]{b8f679}$.75$ & \cellcolor[HTML]{def679}$.60$ & \cellcolor[HTML]{dff679}$.59$ & \cellcolor[HTML]{f6e079}$.41$ & \cellcolor[HTML]{f6e779}$.44$ & \cellcolor[HTML]{f0f679}$.53$ & \cellcolor[HTML]{f6cf79}$.34$ & \cellcolor[HTML]{f6da79}$.39$ & \cellcolor[HTML]{f6df79}$.41$ & \cellcolor[HTML]{f6a779}$.19$ & \cellcolor[HTML]{f6d679}$.37$ & \cellcolor[HTML]{f6a779}$.19$ & \cellcolor[HTML]{f6c379}$.30$ & \cellcolor[HTML]{f6d379}$.36$ & \cellcolor[HTML]{f6e479}$.43$ & \cellcolor[HTML]{f6ee79}$.47$ & \cellcolor[HTML]{f6d279}$.35$ & \cellcolor[HTML]{f4f679}$.51$ \\
PSNR-HVS     & 2006 & PSNR with HVS contrast masking               & $\uparrow$ & \cite{egiazarian2006psnrhvs} & \cellcolor[HTML]{a9f679}$.81$ & \cellcolor[HTML]{acf679}$.80$ & & \cellcolor[HTML]{dcf679}$.60$ & \cellcolor[HTML]{f6ec79}$.46$ & \cellcolor[HTML]{d8f679}$.62$ & \cellcolor[HTML]{b6f679}$.76$ & \cellcolor[HTML]{b1f679}$.77$ & \cellcolor[HTML]{dcf679}$.61$ & \cellcolor[HTML]{f6f679}$.50$ & \cellcolor[HTML]{c6f679}$.69$ & \cellcolor[HTML]{c1f679}$.71$ & \cellcolor[HTML]{d1f679}$.65$ & \cellcolor[HTML]{ccf679}$.67$ & \cellcolor[HTML]{d4f679}$.64$ & \cellcolor[HTML]{f6f579}$.49$ & \cellcolor[HTML]{e3f679}$.58$ & \cellcolor[HTML]{acf679}$.80$ & \cellcolor[HTML]{c5f679}$.70$ & \cellcolor[HTML]{a2f679}$.84$ & \cellcolor[HTML]{c3f679}$.71$ & \cellcolor[HTML]{dcf679}$.60$ & \cellcolor[HTML]{f3f679}$.51$ & \cellcolor[HTML]{bff679}$.72$ & \cellcolor[HTML]{f6ee79}$.47$ & \cellcolor[HTML]{e1f679}$.58$ & \cellcolor[HTML]{d0f679}$.65$ & \cellcolor[HTML]{f6de79}$.40$ & \cellcolor[HTML]{f6f579}$.49$ & \cellcolor[HTML]{f6d479}$.36$ & \cellcolor[HTML]{f6e879}$.44$ & \cellcolor[HTML]{f6e979}$.45$ & \cellcolor[HTML]{d8f679}$.62$ & \cellcolor[HTML]{c2f679}$.71$ & \cellcolor[HTML]{f2f679}$.52$ & \cellcolor[HTML]{bff679}$.72$ \\
mDCTPSNR     & 2009 & Modified DCT-based Perceptual Similarity & $\downarrow$& \cite{richter2009mdct} & \cellcolor[HTML]{d4f679}$.64$ & \cellcolor[HTML]{c1f679}$.71$ & \cellcolor[HTML]{abf679}$.80$ & & \cellcolor[HTML]{f6d979}$.38$ & \cellcolor[HTML]{caf679}$.68$ & \cellcolor[HTML]{cbf679}$.67$ & \cellcolor[HTML]{ccf679}$.67$ & \cellcolor[HTML]{cef679}$.66$ & \cellcolor[HTML]{f6dd79}$.40$ & \cellcolor[HTML]{f2f679}$.52$ & \cellcolor[HTML]{eff679}$.53$ & \cellcolor[HTML]{f6f579}$.50$ & \cellcolor[HTML]{d1f679}$.65$ & \cellcolor[HTML]{edf679}$.54$ & \cellcolor[HTML]{f6ce79}$.34$ & \cellcolor[HTML]{f6db79}$.39$ & \cellcolor[HTML]{d9f679}$.62$ & \cellcolor[HTML]{d2f679}$.64$ & \cellcolor[HTML]{d4f679}$.64$ & \cellcolor[HTML]{f6f279}$.48$ & \cellcolor[HTML]{e0f679}$.59$ & \cellcolor[HTML]{b5f679}$.76$ & \cellcolor[HTML]{caf679}$.68$ & \cellcolor[HTML]{cef679}$.66$ & \cellcolor[HTML]{d8f679}$.62$ & \cellcolor[HTML]{e0f679}$.59$ & \cellcolor[HTML]{f6f379}$.49$ & \cellcolor[HTML]{d3f679}$.64$ & \cellcolor[HTML]{f6d379}$.36$ & \cellcolor[HTML]{d5f679}$.63$ & \cellcolor[HTML]{f6d279}$.36$ & \cellcolor[HTML]{e2f679}$.58$ & \cellcolor[HTML]{c6f679}$.69$ & \cellcolor[HTML]{f6f479}$.49$ & \cellcolor[HTML]{c2f679}$.71$ \\

SSIM         & 2003 & Structural Similarity Index (libvmaf)       & $\uparrow$ & \cite{ssim} & \cellcolor[HTML]{f4f679}$.51$ & \cellcolor[HTML]{edf679}$.54$ & \cellcolor[HTML]{d3f679}$.64$ & \cellcolor[HTML]{edf679}$.54$ & & \cellcolor[HTML]{f6ee79}$.47$ & \cellcolor[HTML]{f6eb79}$.46$ & \cellcolor[HTML]{f6f379}$.49$ & \cellcolor[HTML]{f6e979}$.45$ & \cellcolor[HTML]{aaf679}$.80$ & \cellcolor[HTML]{f6e479}$.43$ & \cellcolor[HTML]{f6e679}$.44$ & \cellcolor[HTML]{f6d779}$.38$ & \cellcolor[HTML]{f6e379}$.42$ & \cellcolor[HTML]{f6d879}$.38$ & \cellcolor[HTML]{f6c079}$.28$ & \cellcolor[HTML]{c0f679}$.72$ & \cellcolor[HTML]{f6f479}$.49$ & \cellcolor[HTML]{f6ea79}$.45$ & \cellcolor[HTML]{f6ea79}$.45$ & \cellcolor[HTML]{f6e179}$.41$ & \cellcolor[HTML]{f0f679}$.53$ & \cellcolor[HTML]{f6d379}$.36$ & \cellcolor[HTML]{f6f479}$.49$ & \cellcolor[HTML]{f6c479}$.30$ & \cellcolor[HTML]{f6cf79}$.34$ & \cellcolor[HTML]{f6d979}$.38$ & \cellcolor[HTML]{f6c679}$.31$ & \cellcolor[HTML]{f6d079}$.35$ & \cellcolor[HTML]{f6c079}$.28$ & \cellcolor[HTML]{f6c179}$.29$ & \cellcolor[HTML]{f1f679}$.52$ & \cellcolor[HTML]{f6d779}$.38$ & \cellcolor[HTML]{f6dd79}$.40$ & \cellcolor[HTML]{f6cb79}$.33$ & \cellcolor[HTML]{f6e279}$.42$ \\
MS-SSIM      & 2003 & Multi-Scale SSIM (libvmaf)                & $\uparrow$ & \cite{ssim} & \cellcolor[HTML]{d9f679}$.62$ & \cellcolor[HTML]{c5f679}$.70$ & \cellcolor[HTML]{a7f679}$.81$ & \cellcolor[HTML]{9cf679}$.86$ & \cellcolor[HTML]{d9f679}$.62$ & & \cellcolor[HTML]{ccf679}$.67$ & \cellcolor[HTML]{c1f679}$.71$ & \cellcolor[HTML]{b2f679}$.77$ & \cellcolor[HTML]{f6de79}$.40$ & \cellcolor[HTML]{dbf679}$.61$ & \cellcolor[HTML]{d9f679}$.62$ & \cellcolor[HTML]{e8f679}$.56$ & \cellcolor[HTML]{d3f679}$.64$ & \cellcolor[HTML]{d7f679}$.63$ & \cellcolor[HTML]{f6d679}$.37$ & \cellcolor[HTML]{f6da79}$.39$ & \cellcolor[HTML]{d7f679}$.63$ & \cellcolor[HTML]{baf679}$.74$ & \cellcolor[HTML]{daf679}$.61$ & \cellcolor[HTML]{f6f379}$.49$ & \cellcolor[HTML]{e0f679}$.59$ & \cellcolor[HTML]{bef679}$.72$ & \cellcolor[HTML]{b2f679}$.77$ & \cellcolor[HTML]{d7f679}$.63$ & \cellcolor[HTML]{dbf679}$.61$ & \cellcolor[HTML]{f1f679}$.52$ & \cellcolor[HTML]{f6ef79}$.47$ & \cellcolor[HTML]{f3f679}$.52$ & \cellcolor[HTML]{f6c779}$.31$ & \cellcolor[HTML]{f1f679}$.52$ & \cellcolor[HTML]{f6d279}$.35$ & \cellcolor[HTML]{eff679}$.53$ & \cellcolor[HTML]{dff679}$.59$ & \cellcolor[HTML]{f6f279}$.48$ & \cellcolor[HTML]{d2f679}$.65$ \\
SSIMU2  & 2023 & SSIMULACRA v2.1 (XYB MS-SSIM)   & $\uparrow$ & \cite{sneyers2022ssimulacra2} & \cellcolor[HTML]{bbf679}$.74$ & \cellcolor[HTML]{b3f679}$.77$ & \cellcolor[HTML]{8df679}$.92$ & \cellcolor[HTML]{9ef679}$.85$ & \cellcolor[HTML]{d5f679}$.63$ & \cellcolor[HTML]{9ef679}$.85$ & & \cellcolor[HTML]{9cf679}$.86$ & \cellcolor[HTML]{c6f679}$.69$ & \cellcolor[HTML]{f6f479}$.49$ & \cellcolor[HTML]{d0f679}$.65$ & \cellcolor[HTML]{caf679}$.68$ & \cellcolor[HTML]{d8f679}$.62$ & \cellcolor[HTML]{cbf679}$.67$ & \cellcolor[HTML]{d8f679}$.62$ & \cellcolor[HTML]{f6e979}$.45$ & \cellcolor[HTML]{f3f679}$.51$ & \cellcolor[HTML]{b9f679}$.75$ & \cellcolor[HTML]{c3f679}$.71$ & \cellcolor[HTML]{aaf679}$.80$ & \cellcolor[HTML]{daf679}$.61$ & \cellcolor[HTML]{cef679}$.66$ & \cellcolor[HTML]{dbf679}$.61$ & \cellcolor[HTML]{b3f679}$.77$ & \cellcolor[HTML]{f4f679}$.51$ & \cellcolor[HTML]{e6f679}$.56$ & \cellcolor[HTML]{d6f679}$.63$ & \cellcolor[HTML]{f6e979}$.45$ & \cellcolor[HTML]{e7f679}$.56$ & \cellcolor[HTML]{f6d479}$.36$ & \cellcolor[HTML]{f5f679}$.51$ & \cellcolor[HTML]{f6f079}$.48$ & \cellcolor[HTML]{cef679}$.66$ & \cellcolor[HTML]{c5f679}$.70$ & \cellcolor[HTML]{f3f679}$.51$ & \cellcolor[HTML]{c7f679}$.69$ \\
DSSIM  &  2025 & L*a*b* MS-SSIM variant (v3.4.0)  & $\downarrow$ & \cite{dssim} & \cellcolor[HTML]{b2f679}$.77$ & \cellcolor[HTML]{a7f679}$.82$ & \cellcolor[HTML]{8bf679}$.93$ & \cellcolor[HTML]{9ef679}$.85$ & \cellcolor[HTML]{cdf679}$.67$ & \cellcolor[HTML]{95f679}$.89$ & \cellcolor[HTML]{80f679}$.97$ & & \cellcolor[HTML]{c7f679}$.69$ & \cellcolor[HTML]{f6f379}$.49$ & \cellcolor[HTML]{cef679}$.66$ & \cellcolor[HTML]{caf679}$.68$ & \cellcolor[HTML]{dcf679}$.60$ & \cellcolor[HTML]{c9f679}$.68$ & \cellcolor[HTML]{d3f679}$.64$ & \cellcolor[HTML]{f6f279}$.48$ & \cellcolor[HTML]{f4f679}$.51$ & \cellcolor[HTML]{b4f679}$.76$ & \cellcolor[HTML]{b2f679}$.77$ & \cellcolor[HTML]{abf679}$.80$ & \cellcolor[HTML]{d7f679}$.62$ & \cellcolor[HTML]{cdf679}$.66$ & \cellcolor[HTML]{d7f679}$.62$ & \cellcolor[HTML]{aaf679}$.80$ & \cellcolor[HTML]{f0f679}$.52$ & \cellcolor[HTML]{dff679}$.59$ & \cellcolor[HTML]{d9f679}$.62$ & \cellcolor[HTML]{f6e479}$.43$ & \cellcolor[HTML]{edf679}$.54$ & \cellcolor[HTML]{f6d079}$.35$ & \cellcolor[HTML]{f6f379}$.49$ & \cellcolor[HTML]{f6e879}$.44$ & \cellcolor[HTML]{d8f679}$.62$ & \cellcolor[HTML]{cff679}$.66$ & \cellcolor[HTML]{f1f679}$.52$ & \cellcolor[HTML]{c6f679}$.69$ \\
IW-SSIM      & 2011 & Information-Weighted SSIM                    & $\uparrow$ & \cite{wang2011iwssim} & \cellcolor[HTML]{f6f579}$.49$ & \cellcolor[HTML]{e4f679}$.57$ & \cellcolor[HTML]{abf679}$.80$ & \cellcolor[HTML]{a0f679}$.84$ & \cellcolor[HTML]{dcf679}$.60$ & \cellcolor[HTML]{8ef679}$.91$ & \cellcolor[HTML]{99f679}$.87$ & \cellcolor[HTML]{99f679}$.87$ & & \cellcolor[HTML]{f6ed79}$.46$ & \cellcolor[HTML]{cbf679}$.67$ & \cellcolor[HTML]{caf679}$.68$ & \cellcolor[HTML]{e4f679}$.57$ & \cellcolor[HTML]{b6f679}$.76$ & \cellcolor[HTML]{c9f679}$.68$ & \cellcolor[HTML]{f6b579}$.24$ & \cellcolor[HTML]{f6dd79}$.40$ & \cellcolor[HTML]{cff679}$.66$ & \cellcolor[HTML]{dcf679}$.60$ & \cellcolor[HTML]{d2f679}$.64$ & \cellcolor[HTML]{f6e379}$.42$ & \cellcolor[HTML]{d2f679}$.64$ & \cellcolor[HTML]{c8f679}$.68$ & \cellcolor[HTML]{b8f679}$.75$ & \cellcolor[HTML]{d9f679}$.62$ & \cellcolor[HTML]{ddf679}$.60$ & \cellcolor[HTML]{d6f679}$.63$ & \cellcolor[HTML]{d7f679}$.63$ & \cellcolor[HTML]{dbf679}$.61$ & \cellcolor[HTML]{f6d079}$.35$ & \cellcolor[HTML]{daf679}$.61$ & \cellcolor[HTML]{f6d679}$.37$ & \cellcolor[HTML]{d8f679}$.62$ & \cellcolor[HTML]{d7f679}$.62$ & \cellcolor[HTML]{eaf679}$.55$ & \cellcolor[HTML]{d0f679}$.65$ \\
FSIM         & 2011 & Feature Similarity Index                     & $\uparrow$ & \cite{zhang2011fsim} & \cellcolor[HTML]{f6f279}$.48$ & \cellcolor[HTML]{f5f679}$.51$ & \cellcolor[HTML]{c9f679}$.68$ & \cellcolor[HTML]{e7f679}$.56$ & \cellcolor[HTML]{8af679}$.93$ & \cellcolor[HTML]{e9f679}$.55$ & \cellcolor[HTML]{ccf679}$.67$ & \cellcolor[HTML]{cef679}$.66$ & \cellcolor[HTML]{d6f679}$.63$ & & \cellcolor[HTML]{f6f279}$.48$ & \cellcolor[HTML]{f6f579}$.50$ & \cellcolor[HTML]{f6e479}$.43$ & \cellcolor[HTML]{f6f279}$.48$ & \cellcolor[HTML]{f6e179}$.41$ & \cellcolor[HTML]{f6b479}$.24$ & \cellcolor[HTML]{b1f679}$.77$ & \cellcolor[HTML]{e9f679}$.55$ & \cellcolor[HTML]{f6e079}$.41$ & \cellcolor[HTML]{f2f679}$.52$ & \cellcolor[HTML]{f6e479}$.43$ & \cellcolor[HTML]{ebf679}$.55$ & \cellcolor[HTML]{f6cc79}$.33$ & \cellcolor[HTML]{f6f079}$.48$ & \cellcolor[HTML]{f6bc79}$.27$ & \cellcolor[HTML]{f6cc79}$.33$ & \cellcolor[HTML]{f6ec79}$.46$ & \cellcolor[HTML]{f6d379}$.36$ & \cellcolor[HTML]{f6dc79}$.40$ & \cellcolor[HTML]{f6c679}$.31$ & \cellcolor[HTML]{f6cd79}$.34$ & \cellcolor[HTML]{e8f679}$.56$ & \cellcolor[HTML]{f6e779}$.44$ & \cellcolor[HTML]{f6ec79}$.46$ & \cellcolor[HTML]{f6d579}$.37$ & \cellcolor[HTML]{f6e979}$.45$ \\

VMAF v0     & 2016 & Video Multimethod Assessment Fusion    & $\uparrow$ & \cite{li2016vmaf} & \cellcolor[HTML]{def679}$.60$ & \cellcolor[HTML]{d7f679}$.62$ & \cellcolor[HTML]{97f679}$.88$ & \cellcolor[HTML]{c1f679}$.71$ & \cellcolor[HTML]{def679}$.60$ & \cellcolor[HTML]{abf679}$.80$ & \cellcolor[HTML]{9ff679}$.85$ & \cellcolor[HTML]{9ff679}$.85$ & \cellcolor[HTML]{9cf679}$.86$ & \cellcolor[HTML]{cdf679}$.66$ & & \cellcolor[HTML]{8df679}$.92$ & \cellcolor[HTML]{c1f679}$.71$ & \cellcolor[HTML]{c9f679}$.68$ & \cellcolor[HTML]{9ff679}$.85$ & \cellcolor[HTML]{f6ca79}$.32$ & \cellcolor[HTML]{f6f379}$.49$ & \cellcolor[HTML]{c3f679}$.71$ & \cellcolor[HTML]{e2f679}$.58$ & \cellcolor[HTML]{c6f679}$.69$ & \cellcolor[HTML]{f0f679}$.53$ & \cellcolor[HTML]{e7f679}$.56$ & \cellcolor[HTML]{f6f079}$.48$ & \cellcolor[HTML]{ccf679}$.67$ & \cellcolor[HTML]{f6ec79}$.46$ & \cellcolor[HTML]{ebf679}$.55$ & \cellcolor[HTML]{d9f679}$.62$ & \cellcolor[HTML]{f6f679}$.50$ & \cellcolor[HTML]{f6f579}$.50$ & \cellcolor[HTML]{f6c379}$.30$ & \cellcolor[HTML]{f6ea79}$.45$ & \cellcolor[HTML]{f6de79}$.40$ & \cellcolor[HTML]{dcf679}$.60$ & \cellcolor[HTML]{e1f679}$.59$ & \cellcolor[HTML]{f5f679}$.51$ & \cellcolor[HTML]{d9f679}$.62$ \\
VMAF-neg v0 & 2020 & VMAF with No Enhancement Gain          & $\uparrow$ & \cite{vmaf-neg} & \cellcolor[HTML]{daf679}$.61$ & \cellcolor[HTML]{d5f679}$.63$ & \cellcolor[HTML]{94f679}$.89$ & \cellcolor[HTML]{bef679}$.72$ & \cellcolor[HTML]{dbf679}$.61$ & \cellcolor[HTML]{a9f679}$.81$ & \cellcolor[HTML]{9af679}$.87$ & \cellcolor[HTML]{9cf679}$.86$ & \cellcolor[HTML]{9bf679}$.86$ & \cellcolor[HTML]{caf679}$.68$ & \cellcolor[HTML]{7bf679}$.99$ & & \cellcolor[HTML]{b6f679}$.76$ & \cellcolor[HTML]{c6f679}$.69$ & \cellcolor[HTML]{a7f679}$.82$ & \cellcolor[HTML]{f6cd79}$.34$ & \cellcolor[HTML]{f5f679}$.50$ & \cellcolor[HTML]{c1f679}$.71$ & \cellcolor[HTML]{dff679}$.59$ & \cellcolor[HTML]{c0f679}$.72$ & \cellcolor[HTML]{ebf679}$.55$ & \cellcolor[HTML]{e3f679}$.58$ & \cellcolor[HTML]{f6f279}$.48$ & \cellcolor[HTML]{c8f679}$.68$ & \cellcolor[HTML]{f6ef79}$.47$ & \cellcolor[HTML]{e7f679}$.56$ & \cellcolor[HTML]{d9f679}$.62$ & \cellcolor[HTML]{f5f679}$.51$ & \cellcolor[HTML]{f4f679}$.51$ & \cellcolor[HTML]{f6c279}$.29$ & \cellcolor[HTML]{f6ec79}$.46$ & \cellcolor[HTML]{f6e579}$.43$ & \cellcolor[HTML]{d7f679}$.62$ & \cellcolor[HTML]{daf679}$.61$ & \cellcolor[HTML]{f3f679}$.51$ & \cellcolor[HTML]{d4f679}$.64$ \\
VMAF v1     & 2026 & VMAF v1    & $\uparrow$ & \cite{bampis2026vmafv1} & \cellcolor[HTML]{daf679}$.61$ & \cellcolor[HTML]{d9f679}$.62$ & \cellcolor[HTML]{a2f679}$.84$ & \cellcolor[HTML]{c8f679}$.68$ & \cellcolor[HTML]{ecf679}$.54$ & \cellcolor[HTML]{b7f679}$.75$ & \cellcolor[HTML]{a8f679}$.81$ & \cellcolor[HTML]{acf679}$.80$ & \cellcolor[HTML]{b2f679}$.77$ & \cellcolor[HTML]{dcf679}$.61$ & \cellcolor[HTML]{95f679}$.89$ & \cellcolor[HTML]{8df679}$.92$ & & \cellcolor[HTML]{def679}$.60$ & \cellcolor[HTML]{c5f679}$.70$ & \cellcolor[HTML]{f6d879}$.38$ & \cellcolor[HTML]{f6f179}$.48$ & \cellcolor[HTML]{e2f679}$.58$ & \cellcolor[HTML]{ebf679}$.54$ & \cellcolor[HTML]{cff679}$.66$ & \cellcolor[HTML]{ebf679}$.55$ & \cellcolor[HTML]{f6f679}$.50$ & \cellcolor[HTML]{f6ea79}$.45$ & \cellcolor[HTML]{dff679}$.60$ & \cellcolor[HTML]{f6f179}$.48$ & \cellcolor[HTML]{edf679}$.54$ & \cellcolor[HTML]{e9f679}$.55$ & \cellcolor[HTML]{f6eb79}$.45$ & \cellcolor[HTML]{f1f679}$.52$ & \cellcolor[HTML]{f6bf79}$.28$ & \cellcolor[HTML]{f6f079}$.48$ & \cellcolor[HTML]{f6f679}$.50$ & \cellcolor[HTML]{d9f679}$.62$ & \cellcolor[HTML]{d9f679}$.62$ & \cellcolor[HTML]{f0f679}$.52$ & \cellcolor[HTML]{d7f679}$.63$ \\
VIF         & 2006           & Visual Information Fidelity                  & $\uparrow$ & \cite{sheikh2006vif} & \cellcolor[HTML]{ecf679}$.54$ & \cellcolor[HTML]{dbf679}$.61$ & \cellcolor[HTML]{9ef679}$.85$ & \cellcolor[HTML]{a3f679}$.83$ & \cellcolor[HTML]{e2f679}$.58$ & \cellcolor[HTML]{a6f679}$.82$ & \cellcolor[HTML]{9ef679}$.85$ & \cellcolor[HTML]{9df679}$.86$ & \cellcolor[HTML]{8df679}$.92$ & \cellcolor[HTML]{d0f679}$.65$ & \cellcolor[HTML]{9bf679}$.86$ & \cellcolor[HTML]{99f679}$.87$ & \cellcolor[HTML]{aff679}$.79$ & & \cellcolor[HTML]{cdf679}$.66$ & \cellcolor[HTML]{f6ba79}$.26$ & \cellcolor[HTML]{f6e679}$.43$ & \cellcolor[HTML]{c0f679}$.72$ & \cellcolor[HTML]{e2f679}$.58$ & \cellcolor[HTML]{bff679}$.72$ & \cellcolor[HTML]{f6f079}$.48$ & \cellcolor[HTML]{d3f679}$.64$ & \cellcolor[HTML]{dbf679}$.61$ & \cellcolor[HTML]{bdf679}$.73$ & \cellcolor[HTML]{e2f679}$.58$ & \cellcolor[HTML]{dbf679}$.61$ & \cellcolor[HTML]{cbf679}$.67$ & \cellcolor[HTML]{d8f679}$.62$ & \cellcolor[HTML]{dbf679}$.61$ & \cellcolor[HTML]{f6d679}$.37$ & \cellcolor[HTML]{e1f679}$.58$ & \cellcolor[HTML]{f6d479}$.36$ & \cellcolor[HTML]{d4f679}$.64$ & \cellcolor[HTML]{cdf679}$.66$ & \cellcolor[HTML]{ebf679}$.55$ & \cellcolor[HTML]{caf679}$.68$ \\
ADM2          & 2016 & Adaptive Detail Metric    & $\uparrow$ & \cite{li2016vmaf} & \cellcolor[HTML]{dcf679}$.61$ & \cellcolor[HTML]{d3f679}$.64$ & \cellcolor[HTML]{a4f679}$.83$ & \cellcolor[HTML]{bcf679}$.73$ & \cellcolor[HTML]{ecf679}$.54$ & \cellcolor[HTML]{a7f679}$.82$ & \cellcolor[HTML]{a7f679}$.82$ & \cellcolor[HTML]{a3f679}$.83$ & \cellcolor[HTML]{9af679}$.87$ & \cellcolor[HTML]{e2f679}$.58$ & \cellcolor[HTML]{81f679}$.97$ & \cellcolor[HTML]{84f679}$.95$ & \cellcolor[HTML]{98f679}$.88$ & \cellcolor[HTML]{9df679}$.85$ & & \cellcolor[HTML]{f6c979}$.32$ & \cellcolor[HTML]{f6e379}$.42$ & \cellcolor[HTML]{d8f679}$.62$ & \cellcolor[HTML]{e0f679}$.59$ & \cellcolor[HTML]{d2f679}$.65$ & \cellcolor[HTML]{f6f179}$.48$ & \cellcolor[HTML]{f1f679}$.52$ & \cellcolor[HTML]{f3f679}$.51$ & \cellcolor[HTML]{d1f679}$.65$ & \cellcolor[HTML]{f5f679}$.50$ & \cellcolor[HTML]{e8f679}$.56$ & \cellcolor[HTML]{dff679}$.59$ & \cellcolor[HTML]{f0f679}$.53$ & \cellcolor[HTML]{eef679}$.53$ & \cellcolor[HTML]{f6b879}$.25$ & \cellcolor[HTML]{f6f579}$.49$ & \cellcolor[HTML]{f6d679}$.37$ & \cellcolor[HTML]{def679}$.60$ & \cellcolor[HTML]{e8f679}$.56$ & \cellcolor[HTML]{eff679}$.53$ & \cellcolor[HTML]{dbf679}$.61$ \\
CIEDE2000     & 2001 & Mean CIE $\Delta E_{00}$ color difference   & $\downarrow$ & \cite{sharma2005ciede2000} & \cellcolor[HTML]{96f679}$.88$ & \cellcolor[HTML]{a5f679}$.83$ & \cellcolor[HTML]{caf679}$.68$ & \cellcolor[HTML]{f6f579}$.49$ & \cellcolor[HTML]{f6e279}$.42$ & \cellcolor[HTML]{eef679}$.53$ & \cellcolor[HTML]{d7f679}$.63$ & \cellcolor[HTML]{cbf679}$.67$ & \cellcolor[HTML]{f6d279}$.36$ & \cellcolor[HTML]{f6d179}$.35$ & \cellcolor[HTML]{f6ec79}$.46$ & \cellcolor[HTML]{f6f079}$.48$ & \cellcolor[HTML]{eef679}$.53$ & \cellcolor[HTML]{f6d879}$.38$ & \cellcolor[HTML]{f6eb79}$.45$ & & \cellcolor[HTML]{f6da79}$.39$ & \cellcolor[HTML]{f6d979}$.38$ & \cellcolor[HTML]{e6f679}$.56$ & \cellcolor[HTML]{f6ec79}$.46$ & \cellcolor[HTML]{d9f679}$.62$ & \cellcolor[HTML]{f6c179}$.29$ & \cellcolor[HTML]{f6c779}$.31$ & \cellcolor[HTML]{f6db79}$.39$ & \cellcolor[HTML]{f6bd79}$.27$ & \cellcolor[HTML]{f6cd79}$.34$ & \cellcolor[HTML]{f6c879}$.31$ & \cellcolor[HTML]{f68279}$.04$ & \cellcolor[HTML]{f6b779}$.25$ & \cellcolor[HTML]{f6a879}$.19$ & \cellcolor[HTML]{f6aa79}$.20$ & \cellcolor[HTML]{f6cf79}$.34$ & \cellcolor[HTML]{f6c879}$.31$ & \cellcolor[HTML]{f6da79}$.39$ & \cellcolor[HTML]{f6bf79}$.28$ & \cellcolor[HTML]{f6dd79}$.40$ \\
VSI       & 2014 & Visual Saliency-Induced Index                 & $\uparrow$ & \cite{zhang2014vsi} & \cellcolor[HTML]{d0f679}$.65$ & \cellcolor[HTML]{d8f679}$.62$ & \cellcolor[HTML]{b5f679}$.76$ & \cellcolor[HTML]{e9f679}$.55$ & \cellcolor[HTML]{94f679}$.89$ & \cellcolor[HTML]{ebf679}$.55$ & \cellcolor[HTML]{c5f679}$.70$ & \cellcolor[HTML]{c6f679}$.69$ & \cellcolor[HTML]{e7f679}$.56$ & \cellcolor[HTML]{8af679}$.93$ & \cellcolor[HTML]{cbf679}$.67$ & \cellcolor[HTML]{c6f679}$.69$ & \cellcolor[HTML]{cdf679}$.67$ & \cellcolor[HTML]{dcf679}$.60$ & \cellcolor[HTML]{dff679}$.59$ & \cellcolor[HTML]{eaf679}$.55$ & & \cellcolor[HTML]{e9f679}$.55$ & \cellcolor[HTML]{f6ec79}$.46$ & \cellcolor[HTML]{e7f679}$.56$ & \cellcolor[HTML]{edf679}$.54$ & \cellcolor[HTML]{f6f679}$.50$ & \cellcolor[HTML]{f6c579}$.30$ & \cellcolor[HTML]{f6ef79}$.47$ & \cellcolor[HTML]{f6bb79}$.26$ & \cellcolor[HTML]{f6d379}$.36$ & \cellcolor[HTML]{f6f079}$.47$ & \cellcolor[HTML]{f6bf79}$.28$ & \cellcolor[HTML]{f6da79}$.38$ & \cellcolor[HTML]{f6c979}$.32$ & \cellcolor[HTML]{f6c779}$.31$ & \cellcolor[HTML]{ddf679}$.60$ & \cellcolor[HTML]{f6ec79}$.46$ & \cellcolor[HTML]{f6f679}$.50$ & \cellcolor[HTML]{f6d879}$.38$ & \cellcolor[HTML]{f6f679}$.50$ \\
GMSD       & 2014  & Gradient Magnitude Similarity Deviation      & $\downarrow$ & \cite{xue2014gmsd} & \cellcolor[HTML]{c5f679}$.70$ & \cellcolor[HTML]{bbf679}$.74$ & \cellcolor[HTML]{87f679}$.94$ & \cellcolor[HTML]{a9f679}$.81$ & \cellcolor[HTML]{cbf679}$.67$ & \cellcolor[HTML]{a9f679}$.81$ & \cellcolor[HTML]{8ff679}$.91$ & \cellcolor[HTML]{8df679}$.92$ & \cellcolor[HTML]{a0f679}$.84$ & \cellcolor[HTML]{bcf679}$.73$ & \cellcolor[HTML]{95f679}$.89$ & \cellcolor[HTML]{94f679}$.89$ & \cellcolor[HTML]{b0f679}$.78$ & \cellcolor[HTML]{96f679}$.88$ & \cellcolor[HTML]{a6f679}$.82$ & \cellcolor[HTML]{ecf679}$.54$ & \cellcolor[HTML]{b9f679}$.74$ & & \cellcolor[HTML]{cff679}$.66$ & \cellcolor[HTML]{a5f679}$.82$ & \cellcolor[HTML]{e2f679}$.58$ & \cellcolor[HTML]{cef679}$.66$ & \cellcolor[HTML]{f1f679}$.52$ & \cellcolor[HTML]{bcf679}$.73$ & \cellcolor[HTML]{f6e979}$.45$ & \cellcolor[HTML]{eaf679}$.55$ & \cellcolor[HTML]{cef679}$.66$ & \cellcolor[HTML]{f6e679}$.43$ & \cellcolor[HTML]{f6f479}$.49$ & \cellcolor[HTML]{f6d479}$.36$ & \cellcolor[HTML]{f6e679}$.43$ & \cellcolor[HTML]{f6e179}$.41$ & \cellcolor[HTML]{dbf679}$.61$ & \cellcolor[HTML]{d0f679}$.65$ & \cellcolor[HTML]{f6f079}$.48$ & \cellcolor[HTML]{c9f679}$.68$ \\
NLPD        & 2016           & Normalized Laplacian Pyramid Distance        & $\downarrow$ & \cite{laparra2016nlpd} & \cellcolor[HTML]{9ef679}$.85$ & \cellcolor[HTML]{8ff679}$.91$ & \cellcolor[HTML]{97f679}$.88$ & \cellcolor[HTML]{a3f679}$.83$ & \cellcolor[HTML]{d7f679}$.63$ & \cellcolor[HTML]{90f679}$.91$ & \cellcolor[HTML]{98f679}$.88$ & \cellcolor[HTML]{8bf679}$.93$ & \cellcolor[HTML]{adf679}$.79$ & \cellcolor[HTML]{e3f679}$.58$ & \cellcolor[HTML]{b3f679}$.77$ & \cellcolor[HTML]{b0f679}$.78$ & \cellcolor[HTML]{bbf679}$.74$ & \cellcolor[HTML]{b3f679}$.77$ & \cellcolor[HTML]{b0f679}$.78$ & \cellcolor[HTML]{b6f679}$.76$ & \cellcolor[HTML]{d4f679}$.64$ & \cellcolor[HTML]{a2f679}$.84$ & & \cellcolor[HTML]{c9f679}$.68$ & \cellcolor[HTML]{d7f679}$.63$ & \cellcolor[HTML]{eef679}$.53$ & \cellcolor[HTML]{d9f679}$.62$ & \cellcolor[HTML]{c2f679}$.71$ & \cellcolor[HTML]{f6f679}$.50$ & \cellcolor[HTML]{eff679}$.53$ & \cellcolor[HTML]{f6f679}$.50$ & \cellcolor[HTML]{f6cc79}$.33$ & \cellcolor[HTML]{f6ed79}$.46$ & \cellcolor[HTML]{f6bc79}$.27$ & \cellcolor[HTML]{f6e479}$.43$ & \cellcolor[HTML]{f6da79}$.39$ & \cellcolor[HTML]{f5f679}$.51$ & \cellcolor[HTML]{e3f679}$.58$ & \cellcolor[HTML]{f6e979}$.45$ & \cellcolor[HTML]{d5f679}$.63$ \\
HaarPSI       & 2018         & Haar Wavelet Perceptual Similarity           & $\uparrow$ & \cite{reisenhofer2018haarpsi} & \cellcolor[HTML]{b3f679}$.77$ & \cellcolor[HTML]{aff679}$.79$ & \cellcolor[HTML]{82f679}$.96$ & \cellcolor[HTML]{a4f679}$.83$ & \cellcolor[HTML]{d6f679}$.63$ & \cellcolor[HTML]{abf679}$.80$ & \cellcolor[HTML]{87f679}$.94$ & \cellcolor[HTML]{88f679}$.94$ & \cellcolor[HTML]{a2f679}$.84$ & \cellcolor[HTML]{c4f679}$.70$ & \cellcolor[HTML]{97f679}$.88$ & \cellcolor[HTML]{94f679}$.89$ & \cellcolor[HTML]{a1f679}$.84$ & \cellcolor[HTML]{95f679}$.89$ & \cellcolor[HTML]{a2f679}$.84$ & \cellcolor[HTML]{d6f679}$.63$ & \cellcolor[HTML]{b9f679}$.75$ & \cellcolor[HTML]{83f679}$.96$ & \cellcolor[HTML]{9df679}$.85$ & & \cellcolor[HTML]{d3f679}$.64$ & \cellcolor[HTML]{d5f679}$.63$ & \cellcolor[HTML]{ecf679}$.54$ & \cellcolor[HTML]{c0f679}$.72$ & \cellcolor[HTML]{f6f379}$.49$ & \cellcolor[HTML]{e5f679}$.57$ & \cellcolor[HTML]{c8f679}$.68$ & \cellcolor[HTML]{f6e979}$.45$ & \cellcolor[HTML]{e8f679}$.56$ & \cellcolor[HTML]{f6d679}$.37$ & \cellcolor[HTML]{f6f479}$.49$ & \cellcolor[HTML]{f6ef79}$.47$ & \cellcolor[HTML]{c8f679}$.69$ & \cellcolor[HTML]{c1f679}$.71$ & \cellcolor[HTML]{eef679}$.54$ & \cellcolor[HTML]{c0f679}$.72$ \\
FLIP       & 2020  & \FLIP Perceptual Difference Evaluator           & $\downarrow$ & \cite{andersson2020flip} & \cellcolor[HTML]{a6f679}$.82$ & \cellcolor[HTML]{b0f679}$.78$ & \cellcolor[HTML]{96f679}$.88$ & \cellcolor[HTML]{ccf679}$.67$ & \cellcolor[HTML]{dff679}$.59$ & \cellcolor[HTML]{caf679}$.68$ & \cellcolor[HTML]{aaf679}$.80$ & \cellcolor[HTML]{a8f679}$.81$ & \cellcolor[HTML]{ddf679}$.60$ & \cellcolor[HTML]{dcf679}$.60$ & \cellcolor[HTML]{bff679}$.72$ & \cellcolor[HTML]{baf679}$.74$ & \cellcolor[HTML]{baf679}$.74$ & \cellcolor[HTML]{cff679}$.66$ & \cellcolor[HTML]{cdf679}$.66$ & \cellcolor[HTML]{a8f679}$.81$ & \cellcolor[HTML]{bcf679}$.73$ & \cellcolor[HTML]{b1f679}$.78$ & \cellcolor[HTML]{a8f679}$.81$ & \cellcolor[HTML]{a5f679}$.82$ & & \cellcolor[HTML]{f6ea79}$.45$ & \cellcolor[HTML]{f6db79}$.39$ & \cellcolor[HTML]{e8f679}$.56$ & \cellcolor[HTML]{f6d679}$.37$ & \cellcolor[HTML]{f6ee79}$.47$ & \cellcolor[HTML]{f6f579}$.49$ & \cellcolor[HTML]{f6b679}$.25$ & \cellcolor[HTML]{f6db79}$.39$ & \cellcolor[HTML]{f6c579}$.31$ & \cellcolor[HTML]{f6d179}$.35$ & \cellcolor[HTML]{f6ed79}$.46$ & \cellcolor[HTML]{f6f179}$.48$ & \cellcolor[HTML]{def679}$.60$ & \cellcolor[HTML]{f6e979}$.45$ & \cellcolor[HTML]{e3f679}$.58$ \\
HDR-VDP-2  & 2011  & HDR Visual Difference Predictor v2                  & $\uparrow$ & \cite{mantiuk2011hdrvdp2} & \cellcolor[HTML]{eff679}$.53$ & \cellcolor[HTML]{e2f679}$.58$ & \cellcolor[HTML]{abf679}$.80$ & \cellcolor[HTML]{b0f679}$.78$ & \cellcolor[HTML]{bff679}$.72$ & \cellcolor[HTML]{aef679}$.79$ & \cellcolor[HTML]{9ef679}$.85$ & \cellcolor[HTML]{9df679}$.86$ & \cellcolor[HTML]{a1f679}$.84$ & \cellcolor[HTML]{bbf679}$.74$ & \cellcolor[HTML]{b4f679}$.76$ & \cellcolor[HTML]{b1f679}$.78$ & \cellcolor[HTML]{c5f679}$.70$ & \cellcolor[HTML]{a2f679}$.84$ & \cellcolor[HTML]{bff679}$.72$ & \cellcolor[HTML]{f6e379}$.42$ & \cellcolor[HTML]{c6f679}$.69$ & \cellcolor[HTML]{a0f679}$.84$ & \cellcolor[HTML]{bdf679}$.73$ & \cellcolor[HTML]{a5f679}$.83$ & \cellcolor[HTML]{d5f679}$.63$ & & \cellcolor[HTML]{e1f679}$.59$ & \cellcolor[HTML]{c8f679}$.69$ & \cellcolor[HTML]{f6f379}$.49$ & \cellcolor[HTML]{ecf679}$.54$ & \cellcolor[HTML]{d8f679}$.62$ & \cellcolor[HTML]{f6ec79}$.46$ & \cellcolor[HTML]{f5f679}$.51$ & \cellcolor[HTML]{f6e179}$.41$ & \cellcolor[HTML]{f6f179}$.48$ & \cellcolor[HTML]{f6e679}$.43$ & \cellcolor[HTML]{edf679}$.54$ & \cellcolor[HTML]{daf679}$.61$ & \cellcolor[HTML]{f6e779}$.44$ & \cellcolor[HTML]{dbf679}$.61$ \\
HDR-VDP-3   & 2023 & HDR Visual Difference Predictor v3                  & $\uparrow$ & \cite{mantiuk2023hdrvdp3} & \cellcolor[HTML]{eff679}$.53$ & \cellcolor[HTML]{d9f679}$.62$ & \cellcolor[HTML]{c4f679}$.70$ & \cellcolor[HTML]{90f679}$.91$ & \cellcolor[HTML]{f6f679}$.50$ & \cellcolor[HTML]{93f679}$.89$ & \cellcolor[HTML]{acf679}$.80$ & \cellcolor[HTML]{a8f679}$.81$ & \cellcolor[HTML]{9cf679}$.86$ & \cellcolor[HTML]{f6f079}$.47$ & \cellcolor[HTML]{cef679}$.66$ & \cellcolor[HTML]{ccf679}$.67$ & \cellcolor[HTML]{d5f679}$.63$ & \cellcolor[HTML]{acf679}$.80$ & \cellcolor[HTML]{c3f679}$.70$ & \cellcolor[HTML]{f6ea79}$.45$ & \cellcolor[HTML]{f6e779}$.44$ & \cellcolor[HTML]{c3f679}$.70$ & \cellcolor[HTML]{a9f679}$.81$ & \cellcolor[HTML]{bef679}$.73$ & \cellcolor[HTML]{e9f679}$.55$ & \cellcolor[HTML]{b0f679}$.78$ & & \cellcolor[HTML]{cff679}$.66$ & \cellcolor[HTML]{c5f679}$.70$ & \cellcolor[HTML]{def679}$.60$ & \cellcolor[HTML]{eff679}$.53$ & \cellcolor[HTML]{f6f479}$.49$ & \cellcolor[HTML]{dff679}$.59$ & \cellcolor[HTML]{f6d779}$.38$ & \cellcolor[HTML]{d8f679}$.62$ & \cellcolor[HTML]{f6c479}$.30$ & \cellcolor[HTML]{f4f679}$.51$ & \cellcolor[HTML]{ddf679}$.60$ & \cellcolor[HTML]{f6ec79}$.46$ & \cellcolor[HTML]{d9f679}$.62$ \\
CVVDP     & 2023   & Color Video Visual Difference Predictor                                & $\uparrow$ & \cite{mantiuk2023cvvdp} & \cellcolor[HTML]{cef679}$.66$ & \cellcolor[HTML]{bff679}$.72$ & \cellcolor[HTML]{93f679}$.89$ & \cellcolor[HTML]{9cf679}$.86$ & \cellcolor[HTML]{cef679}$.66$ & \cellcolor[HTML]{8af679}$.93$ & \cellcolor[HTML]{8cf679}$.92$ & \cellcolor[HTML]{86f679}$.95$ & \cellcolor[HTML]{8ef679}$.91$ & \cellcolor[HTML]{d3f679}$.64$ & \cellcolor[HTML]{9cf679}$.86$ & \cellcolor[HTML]{99f679}$.87$ & \cellcolor[HTML]{adf679}$.79$ & \cellcolor[HTML]{93f679}$.90$ & \cellcolor[HTML]{a1f679}$.84$ & \cellcolor[HTML]{e7f679}$.56$ & \cellcolor[HTML]{d2f679}$.65$ & \cellcolor[HTML]{92f679}$.90$ & \cellcolor[HTML]{94f679}$.89$ & \cellcolor[HTML]{94f679}$.89$ & \cellcolor[HTML]{b6f679}$.76$ & \cellcolor[HTML]{99f679}$.87$ & \cellcolor[HTML]{9ff679}$.85$ & & \cellcolor[HTML]{e3f679}$.58$ & \cellcolor[HTML]{d3f679}$.64$ & \cellcolor[HTML]{dbf679}$.61$ & \cellcolor[HTML]{f6f479}$.49$ & \cellcolor[HTML]{eef679}$.53$ & \cellcolor[HTML]{f6d379}$.36$ & \cellcolor[HTML]{f6f679}$.50$ & \cellcolor[HTML]{f6dd79}$.40$ & \cellcolor[HTML]{dff679}$.59$ & \cellcolor[HTML]{d1f679}$.65$ & \cellcolor[HTML]{f5f679}$.50$ & \cellcolor[HTML]{c6f679}$.69$ \\

\midrule

LPIPS-alex        & 2018     & Learned Perceptual Image Patch Similarity      & $\downarrow$ & \cite{zhang2018lpips} & \cellcolor[HTML]{f6e879}$.44$ & \cellcolor[HTML]{f6f579}$.49$ & \cellcolor[HTML]{d1f679}$.65$ & \cellcolor[HTML]{a1f679}$.84$ & \cellcolor[HTML]{f6e479}$.42$ & \cellcolor[HTML]{a8f679}$.81$ & \cellcolor[HTML]{c5f679}$.70$ & \cellcolor[HTML]{c1f679}$.71$ & \cellcolor[HTML]{abf679}$.80$ & \cellcolor[HTML]{f6db79}$.39$ & \cellcolor[HTML]{d5f679}$.63$ & \cellcolor[HTML]{d1f679}$.65$ & \cellcolor[HTML]{cff679}$.66$ & \cellcolor[HTML]{b3f679}$.77$ & \cellcolor[HTML]{c7f679}$.69$ & \cellcolor[HTML]{f6de79}$.40$ & \cellcolor[HTML]{f6da79}$.39$ & \cellcolor[HTML]{d8f679}$.62$ & \cellcolor[HTML]{c7f679}$.69$ & \cellcolor[HTML]{ccf679}$.67$ & \cellcolor[HTML]{edf679}$.54$ & \cellcolor[HTML]{cbf679}$.67$ & \cellcolor[HTML]{99f679}$.87$ & \cellcolor[HTML]{b3f679}$.77$ & & \cellcolor[HTML]{c6f679}$.69$ & \cellcolor[HTML]{eff679}$.53$ & \cellcolor[HTML]{e4f679}$.57$ & \cellcolor[HTML]{cef679}$.66$ & \cellcolor[HTML]{f6cf79}$.34$ & \cellcolor[HTML]{c6f679}$.69$ & \cellcolor[HTML]{f6c579}$.30$ & \cellcolor[HTML]{eef679}$.53$ & \cellcolor[HTML]{dcf679}$.61$ & \cellcolor[HTML]{f5f679}$.51$ & \cellcolor[HTML]{d7f679}$.63$ \\
STLPIPS     & 2022 & Shift-Tolerant LPIPS (AlexNet)                      & $\downarrow$ & \cite{ghildyal2022stlpips} & \cellcolor[HTML]{ebf679}$.55$ & \cellcolor[HTML]{e8f679}$.56$ & \cellcolor[HTML]{b1f679}$.78$ & \cellcolor[HTML]{a6f679}$.82$ & \cellcolor[HTML]{f6f379}$.49$ & \cellcolor[HTML]{abf679}$.80$ & \cellcolor[HTML]{b5f679}$.76$ & \cellcolor[HTML]{aef679}$.79$ & \cellcolor[HTML]{adf679}$.79$ & \cellcolor[HTML]{f6ef79}$.47$ & \cellcolor[HTML]{bbf679}$.74$ & \cellcolor[HTML]{b7f679}$.75$ & \cellcolor[HTML]{bdf679}$.73$ & \cellcolor[HTML]{abf679}$.80$ & \cellcolor[HTML]{b8f679}$.75$ & \cellcolor[HTML]{f6f479}$.49$ & \cellcolor[HTML]{f4f679}$.51$ & \cellcolor[HTML]{b9f679}$.74$ & \cellcolor[HTML]{bff679}$.72$ & \cellcolor[HTML]{b4f679}$.77$ & \cellcolor[HTML]{cff679}$.66$ & \cellcolor[HTML]{bbf679}$.74$ & \cellcolor[HTML]{adf679}$.79$ & \cellcolor[HTML]{a4f679}$.83$ & \cellcolor[HTML]{99f679}$.87$ & & \cellcolor[HTML]{daf679}$.61$ & \cellcolor[HTML]{f3f679}$.51$ & \cellcolor[HTML]{e1f679}$.58$ & \cellcolor[HTML]{f6d179}$.35$ & \cellcolor[HTML]{e1f679}$.59$ & \cellcolor[HTML]{f6c979}$.32$ & \cellcolor[HTML]{e7f679}$.56$ & \cellcolor[HTML]{d3f679}$.64$ & \cellcolor[HTML]{f3f679}$.51$ & \cellcolor[HTML]{c4f679}$.70$ \\
PieAPP     & 2018  & Perceptual image-error Assessment 
& $\downarrow$ & \cite{prashnani2018pieapp} & \cellcolor[HTML]{e4f679}$.58$ & \cellcolor[HTML]{e4f679}$.57$ & \cellcolor[HTML]{a1f679}$.84$ & \cellcolor[HTML]{aff679}$.79$ & \cellcolor[HTML]{ebf679}$.55$ & \cellcolor[HTML]{c2f679}$.71$ & \cellcolor[HTML]{a7f679}$.82$ & \cellcolor[HTML]{a9f679}$.81$ & \cellcolor[HTML]{a8f679}$.81$ & \cellcolor[HTML]{d3f679}$.64$ & \cellcolor[HTML]{a8f679}$.81$ & \cellcolor[HTML]{a9f679}$.81$ & \cellcolor[HTML]{baf679}$.74$ & \cellcolor[HTML]{9ff679}$.85$ & \cellcolor[HTML]{aff679}$.78$ & \cellcolor[HTML]{f6e979}$.45$ & \cellcolor[HTML]{cff679}$.66$ & \cellcolor[HTML]{9ff679}$.85$ & \cellcolor[HTML]{c8f679}$.68$ & \cellcolor[HTML]{9bf679}$.87$ & \cellcolor[HTML]{caf679}$.68$ & \cellcolor[HTML]{a8f679}$.81$ & \cellcolor[HTML]{c0f679}$.72$ & \cellcolor[HTML]{abf679}$.80$ & \cellcolor[HTML]{c0f679}$.72$ & \cellcolor[HTML]{aaf679}$.80$ & & \cellcolor[HTML]{f1f679}$.52$ & \cellcolor[HTML]{dff679}$.59$ & \cellcolor[HTML]{f6ea79}$.45$ & \cellcolor[HTML]{e1f679}$.59$ & \cellcolor[HTML]{f6df79}$.41$ & \cellcolor[HTML]{d3f679}$.64$ & \cellcolor[HTML]{c0f679}$.72$ & \cellcolor[HTML]{eaf679}$.55$ & \cellcolor[HTML]{c5f679}$.70$ \\
WaDIQaM    & 2018  & Weighted Average Deep IQA Metric                    & $\uparrow$ & \cite{bosse2018wadiqam} & \cellcolor[HTML]{f6af79}$.22$ & \cellcolor[HTML]{f6c079}$.28$ & \cellcolor[HTML]{e2f679}$.58$ & \cellcolor[HTML]{caf679}$.68$ & \cellcolor[HTML]{f6e979}$.45$ & \cellcolor[HTML]{cff679}$.66$ & \cellcolor[HTML]{d4f679}$.64$ & \cellcolor[HTML]{daf679}$.61$ & \cellcolor[HTML]{a6f679}$.82$ & \cellcolor[HTML]{f3f679}$.51$ & \cellcolor[HTML]{c6f679}$.69$ & \cellcolor[HTML]{c4f679}$.70$ & \cellcolor[HTML]{d3f679}$.64$ & \cellcolor[HTML]{a7f679}$.82$ & \cellcolor[HTML]{bff679}$.72$ & \cellcolor[HTML]{f68779}$.06$ & \cellcolor[HTML]{f6e079}$.41$ & \cellcolor[HTML]{d8f679}$.62$ & \cellcolor[HTML]{f6f479}$.49$ & \cellcolor[HTML]{d4f679}$.64$ & \cellcolor[HTML]{f6d579}$.37$ & \cellcolor[HTML]{d1f679}$.65$ & \cellcolor[HTML]{cbf679}$.67$ & \cellcolor[HTML]{c7f679}$.69$ & \cellcolor[HTML]{b4f679}$.76$ & \cellcolor[HTML]{c3f679}$.71$ & \cellcolor[HTML]{c0f679}$.72$ & & \cellcolor[HTML]{def679}$.60$ & \cellcolor[HTML]{f6c879}$.32$ & \cellcolor[HTML]{e0f679}$.59$ & \cellcolor[HTML]{f6c179}$.29$ & \cellcolor[HTML]{e7f679}$.56$ & \cellcolor[HTML]{f6f479}$.49$ & \cellcolor[HTML]{f6f579}$.49$ & \cellcolor[HTML]{f6f679}$.50$ \\
DISTS      & 2022  & Deep Image Structure and Texture Similarity  & $\downarrow$ & \cite{ding2022dists} & \cellcolor[HTML]{f6f479}$.49$ & \cellcolor[HTML]{eff679}$.53$ & \cellcolor[HTML]{c9f679}$.68$ & \cellcolor[HTML]{a4f679}$.83$ & \cellcolor[HTML]{f6f579}$.50$ & \cellcolor[HTML]{c5f679}$.70$ & \cellcolor[HTML]{b7f679}$.75$ & \cellcolor[HTML]{bdf679}$.73$ & \cellcolor[HTML]{aef679}$.79$ & \cellcolor[HTML]{e8f679}$.56$ & \cellcolor[HTML]{c7f679}$.69$ & \cellcolor[HTML]{c4f679}$.70$ & \cellcolor[HTML]{c1f679}$.71$ & \cellcolor[HTML]{acf679}$.79$ & \cellcolor[HTML]{bcf679}$.73$ & \cellcolor[HTML]{f6d479}$.36$ & \cellcolor[HTML]{ebf679}$.55$ & \cellcolor[HTML]{caf679}$.68$ & \cellcolor[HTML]{d3f679}$.64$ & \cellcolor[HTML]{b7f679}$.75$ & \cellcolor[HTML]{e7f679}$.56$ & \cellcolor[HTML]{c6f679}$.69$ & \cellcolor[HTML]{b1f679}$.78$ & \cellcolor[HTML]{bef679}$.72$ & \cellcolor[HTML]{a1f679}$.84$ & \cellcolor[HTML]{b0f679}$.78$ & \cellcolor[HTML]{adf679}$.79$ & \cellcolor[HTML]{acf679}$.80$ & & \cellcolor[HTML]{f6d479}$.36$ & \cellcolor[HTML]{b0f679}$.78$ & \cellcolor[HTML]{f6e479}$.43$ & \cellcolor[HTML]{cff679}$.66$ & \cellcolor[HTML]{d4f679}$.64$ & \cellcolor[HTML]{dbf679}$.61$ & \cellcolor[HTML]{d4f679}$.64$ \\
AHIQ         & 2022  & Attentive Hybrid Image Quality               & $\uparrow$ & \cite{lao2022ahiq} & \cellcolor[HTML]{f6c379}$.30$ & \cellcolor[HTML]{f6bc79}$.27$ & \cellcolor[HTML]{f1f679}$.52$ & \cellcolor[HTML]{f2f679}$.52$ & \cellcolor[HTML]{f6e179}$.42$ & \cellcolor[HTML]{f6ea79}$.45$ & \cellcolor[HTML]{f1f679}$.52$ & \cellcolor[HTML]{f6f679}$.50$ & \cellcolor[HTML]{f5f679}$.51$ & \cellcolor[HTML]{f6ea79}$.45$ & \cellcolor[HTML]{f6e679}$.44$ & \cellcolor[HTML]{f6e479}$.43$ & \cellcolor[HTML]{f6df79}$.41$ & \cellcolor[HTML]{edf679}$.54$ & \cellcolor[HTML]{f6d679}$.37$ & \cellcolor[HTML]{f6bd79}$.27$ & \cellcolor[HTML]{f6ee79}$.47$ & \cellcolor[HTML]{f2f679}$.52$ & \cellcolor[HTML]{f6da79}$.39$ & \cellcolor[HTML]{eff679}$.53$ & \cellcolor[HTML]{f6e879}$.44$ & \cellcolor[HTML]{e0f679}$.59$ & \cellcolor[HTML]{eef679}$.53$ & \cellcolor[HTML]{f2f679}$.52$ & \cellcolor[HTML]{f6f679}$.50$ & \cellcolor[HTML]{f3f679}$.51$ & \cellcolor[HTML]{d5f679}$.63$ & \cellcolor[HTML]{f6ed79}$.46$ & \cellcolor[HTML]{f0f679}$.52$ & & \cellcolor[HTML]{f6d479}$.36$ & \cellcolor[HTML]{f6be79}$.28$ & \cellcolor[HTML]{f6d579}$.37$ & \cellcolor[HTML]{f6e779}$.44$ & \cellcolor[HTML]{f6c979}$.32$ & \cellcolor[HTML]{f6e379}$.42$ \\
DeepDC     & 2023 & Deep Distance Correlation                     & $\downarrow$ & \cite{zhu2023DeepDC} & \cellcolor[HTML]{f6db79}$.39$ & \cellcolor[HTML]{f6e679}$.44$ & \cellcolor[HTML]{d8f679}$.62$ & \cellcolor[HTML]{a7f679}$.82$ & \cellcolor[HTML]{f6e279}$.42$ & \cellcolor[HTML]{c5f679}$.70$ & \cellcolor[HTML]{c7f679}$.69$ & \cellcolor[HTML]{cbf679}$.67$ & \cellcolor[HTML]{adf679}$.79$ & \cellcolor[HTML]{f6f279}$.48$ & \cellcolor[HTML]{d5f679}$.63$ & \cellcolor[HTML]{d3f679}$.64$ & \cellcolor[HTML]{cef679}$.66$ & \cellcolor[HTML]{b2f679}$.77$ & \cellcolor[HTML]{c8f679}$.68$ & \cellcolor[HTML]{f6c279}$.29$ & \cellcolor[HTML]{f6ea79}$.45$ & \cellcolor[HTML]{dbf679}$.61$ & \cellcolor[HTML]{def679}$.60$ & \cellcolor[HTML]{caf679}$.68$ & \cellcolor[HTML]{f5f679}$.51$ & \cellcolor[HTML]{cef679}$.66$ & \cellcolor[HTML]{aaf679}$.80$ & \cellcolor[HTML]{c7f679}$.69$ & \cellcolor[HTML]{9bf679}$.86$ & \cellcolor[HTML]{b1f679}$.78$ & \cellcolor[HTML]{aff679}$.78$ & \cellcolor[HTML]{aef679}$.79$ & \cellcolor[HTML]{8af679}$.93$ & \cellcolor[HTML]{f0f679}$.52$ & & \cellcolor[HTML]{f6d579}$.37$ & \cellcolor[HTML]{e1f679}$.58$ & \cellcolor[HTML]{d7f679}$.62$ & \cellcolor[HTML]{e1f679}$.58$ & \cellcolor[HTML]{d7f679}$.63$ \\
DreamSim  & 2023 & Perceptual similarity via ensemble features   & $\downarrow$ & \cite{fu2023dreamsim} & \cellcolor[HTML]{f0f679}$.52$ & \cellcolor[HTML]{f2f679}$.52$ & \cellcolor[HTML]{d5f679}$.63$ & \cellcolor[HTML]{f4f679}$.51$ & \cellcolor[HTML]{c0f679}$.72$ & \cellcolor[HTML]{f4f679}$.51$ & \cellcolor[HTML]{cef679}$.66$ & \cellcolor[HTML]{d7f679}$.62$ & \cellcolor[HTML]{eef679}$.53$ & \cellcolor[HTML]{b5f679}$.76$ & \cellcolor[HTML]{e3f679}$.58$ & \cellcolor[HTML]{daf679}$.61$ & \cellcolor[HTML]{c6f679}$.69$ & \cellcolor[HTML]{f1f679}$.52$ & \cellcolor[HTML]{edf679}$.54$ & \cellcolor[HTML]{f6f479}$.49$ & \cellcolor[HTML]{abf679}$.80$ & \cellcolor[HTML]{e0f679}$.59$ & \cellcolor[HTML]{e9f679}$.55$ & \cellcolor[HTML]{d0f679}$.65$ & \cellcolor[HTML]{d1f679}$.65$ & \cellcolor[HTML]{daf679}$.61$ & \cellcolor[HTML]{f6e679}$.43$ & \cellcolor[HTML]{e5f679}$.57$ & \cellcolor[HTML]{f6e779}$.44$ & \cellcolor[HTML]{f6ee79}$.46$ & \cellcolor[HTML]{e2f679}$.58$ & \cellcolor[HTML]{f6e479}$.43$ & \cellcolor[HTML]{def679}$.60$ & \cellcolor[HTML]{f6df79}$.41$ & \cellcolor[HTML]{f1f679}$.52$ & & \cellcolor[HTML]{f6f279}$.48$ & \cellcolor[HTML]{f6ee79}$.47$ & \cellcolor[HTML]{f6de79}$.40$ & \cellcolor[HTML]{f6e579}$.43$ \\
TOPIQ       & 2024   & Top-Down Image Quality (ResNet-50)           & $\uparrow$ & \cite{chen2024topiq} & \cellcolor[HTML]{dff679}$.59$ & \cellcolor[HTML]{dbf679}$.61$ & \cellcolor[HTML]{a6f679}$.82$ & \cellcolor[HTML]{acf679}$.79$ & \cellcolor[HTML]{ecf679}$.54$ & \cellcolor[HTML]{bdf679}$.73$ & \cellcolor[HTML]{9ef679}$.85$ & \cellcolor[HTML]{a6f679}$.82$ & \cellcolor[HTML]{a7f679}$.82$ & \cellcolor[HTML]{d8f679}$.62$ & \cellcolor[HTML]{a9f679}$.81$ & \cellcolor[HTML]{a4f679}$.83$ & \cellcolor[HTML]{a6f679}$.82$ & \cellcolor[HTML]{a4f679}$.83$ & \cellcolor[HTML]{aaf679}$.80$ & \cellcolor[HTML]{f6eb79}$.45$ & \cellcolor[HTML]{d2f679}$.65$ & \cellcolor[HTML]{a8f679}$.81$ & \cellcolor[HTML]{c3f679}$.71$ & \cellcolor[HTML]{98f679}$.88$ & \cellcolor[HTML]{ccf679}$.67$ & \cellcolor[HTML]{baf679}$.74$ & \cellcolor[HTML]{c2f679}$.71$ & \cellcolor[HTML]{acf679}$.80$ & \cellcolor[HTML]{bcf679}$.73$ & \cellcolor[HTML]{b2f679}$.77$ & \cellcolor[HTML]{a1f679}$.84$ & \cellcolor[HTML]{b6f679}$.76$ & \cellcolor[HTML]{9df679}$.86$ & \cellcolor[HTML]{f0f679}$.53$ & \cellcolor[HTML]{adf679}$.79$ & \cellcolor[HTML]{cbf679}$.67$ & & \cellcolor[HTML]{cdf679}$.67$ & \cellcolor[HTML]{dff679}$.59$ & \cellcolor[HTML]{caf679}$.68$ \\
WD          & 2024   & Wasserstein Distortion (VGG-16, $\sigma = 2^4$)              & $\downarrow$ & \cite{qiu2024wd} & \cellcolor[HTML]{d0f679}$.65$ & \cellcolor[HTML]{d1f679}$.65$ & \cellcolor[HTML]{96f679}$.88$ & \cellcolor[HTML]{98f679}$.88$ & \cellcolor[HTML]{e5f679}$.57$ & \cellcolor[HTML]{b0f679}$.78$ & \cellcolor[HTML]{98f679}$.88$ & \cellcolor[HTML]{a0f679}$.84$ & \cellcolor[HTML]{a8f679}$.81$ & \cellcolor[HTML]{d3f679}$.64$ & \cellcolor[HTML]{aff679}$.78$ & \cellcolor[HTML]{a9f679}$.81$ & \cellcolor[HTML]{a9f679}$.81$ & \cellcolor[HTML]{a0f679}$.85$ & \cellcolor[HTML]{b6f679}$.75$ & \cellcolor[HTML]{eaf679}$.55$ & \cellcolor[HTML]{c7f679}$.69$ & \cellcolor[HTML]{a1f679}$.84$ & \cellcolor[HTML]{b2f679}$.77$ & \cellcolor[HTML]{95f679}$.89$ & \cellcolor[HTML]{aef679}$.79$ & \cellcolor[HTML]{a9f679}$.81$ & \cellcolor[HTML]{adf679}$.79$ & \cellcolor[HTML]{a3f679}$.83$ & \cellcolor[HTML]{acf679}$.79$ & \cellcolor[HTML]{a3f679}$.83$ & \cellcolor[HTML]{95f679}$.89$ & \cellcolor[HTML]{c8f679}$.68$ & \cellcolor[HTML]{a4f679}$.83$ & \cellcolor[HTML]{d9f679}$.62$ & \cellcolor[HTML]{a7f679}$.82$ & \cellcolor[HTML]{d1f679}$.65$ & \cellcolor[HTML]{9bf679}$.86$ & & \cellcolor[HTML]{ebf679}$.54$ & \cellcolor[HTML]{aaf679}$.81$ \\
A-FINE     &  2025  & Adaptive Fidelity-Naturalness Evaluator   & $\downarrow$ & \cite{chen2025toward} & \cellcolor[HTML]{f6f579}$.49$ & \cellcolor[HTML]{f6f679}$.50$ & \cellcolor[HTML]{c1f679}$.71$ & \cellcolor[HTML]{c8f679}$.68$ & \cellcolor[HTML]{f6f079}$.48$ & \cellcolor[HTML]{ccf679}$.67$ & \cellcolor[HTML]{c3f679}$.70$ & \cellcolor[HTML]{c1f679}$.71$ & \cellcolor[HTML]{b8f679}$.75$ & \cellcolor[HTML]{eff679}$.53$ & \cellcolor[HTML]{c2f679}$.71$ & \cellcolor[HTML]{c1f679}$.71$ & \cellcolor[HTML]{bff679}$.72$ & \cellcolor[HTML]{b8f679}$.75$ & \cellcolor[HTML]{bcf679}$.73$ & \cellcolor[HTML]{f6df79}$.41$ & \cellcolor[HTML]{ebf679}$.54$ & \cellcolor[HTML]{ccf679}$.67$ & \cellcolor[HTML]{d7f679}$.62$ & \cellcolor[HTML]{bcf679}$.73$ & \cellcolor[HTML]{d6f679}$.63$ & \cellcolor[HTML]{d7f679}$.63$ & \cellcolor[HTML]{d3f679}$.64$ & \cellcolor[HTML]{c4f679}$.70$ & \cellcolor[HTML]{c6f679}$.69$ & \cellcolor[HTML]{c1f679}$.71$ & \cellcolor[HTML]{b9f679}$.75$ & \cellcolor[HTML]{c6f679}$.69$ & \cellcolor[HTML]{a9f679}$.81$ & \cellcolor[HTML]{f6f079}$.47$ & \cellcolor[HTML]{aff679}$.78$ & \cellcolor[HTML]{e6f679}$.57$ & \cellcolor[HTML]{acf679}$.80$ & \cellcolor[HTML]{b8f679}$.75$ & & \cellcolor[HTML]{e4f679}$.57$ \\
DMM       & 2025 & Debiased Mapping based quality Measure    & $\uparrow$ & \cite{chen2025dmm} & \cellcolor[HTML]{c8f679}$.68$ & \cellcolor[HTML]{c5f679}$.70$ & \cellcolor[HTML]{94f679}$.89$ & \cellcolor[HTML]{95f679}$.89$ & \cellcolor[HTML]{dff679}$.59$ & \cellcolor[HTML]{a3f679}$.83$ & \cellcolor[HTML]{9af679}$.87$ & \cellcolor[HTML]{99f679}$.87$ & \cellcolor[HTML]{a2f679}$.83$ & \cellcolor[HTML]{d8f679}$.62$ & \cellcolor[HTML]{a8f679}$.81$ & \cellcolor[HTML]{a5f679}$.83$ & \cellcolor[HTML]{a8f679}$.81$ & \cellcolor[HTML]{9ef679}$.85$ & \cellcolor[HTML]{aaf679}$.80$ & \cellcolor[HTML]{e5f679}$.57$ & \cellcolor[HTML]{c9f679}$.68$ & \cellcolor[HTML]{9cf679}$.86$ & \cellcolor[HTML]{a6f679}$.82$ & \cellcolor[HTML]{95f679}$.89$ & \cellcolor[HTML]{b2f679}$.77$ & \cellcolor[HTML]{aaf679}$.80$ & \cellcolor[HTML]{a9f679}$.81$ & \cellcolor[HTML]{9af679}$.87$ & \cellcolor[HTML]{a7f679}$.81$ & \cellcolor[HTML]{97f679}$.88$ & \cellcolor[HTML]{98f679}$.88$ & \cellcolor[HTML]{c7f679}$.69$ & \cellcolor[HTML]{a4f679}$.83$ & \cellcolor[HTML]{def679}$.60$ & \cellcolor[HTML]{a7f679}$.82$ & \cellcolor[HTML]{dcf679}$.61$ & \cellcolor[HTML]{98f679}$.88$ & \cellcolor[HTML]{86f679}$.95$ & \cellcolor[HTML]{b0f679}$.78$ & \\
\bottomrule
\end{tabular}
}

\end{table*}

To illustrate the range and granularity of the 20 distortion levels of the AIC2026 dataset, Fig.~\historef{} compares the CVVDP image quality score distributions for the distortion levels in TID2013~\cite{ponomarenko2015tid2013}, JPEG AIC-3~\cite{testolina2023jpeg,jenadeleh2025subjective}, and AIC2026 datasets. For each distortion level in each dataset, CVVDP was computed for all distorted images, and the resulting score distributions and median values are shown. The CVVDP scores, mapped to JND units using Equation \ref{CVVDPtoJND}, are also shown on the x-axis.

The AIC2026 dataset is intended to cover the range from  good quality to lossless compression. This range corresponds to distortions ranging from 0 to approximately 3.0 JND \cite{29170-3-AIC3}. 
As shown in Fig.~\historef{}, only the two lowest distortion levels in TID2013 have median JND-mapped CVVDP scores within this range, at approximately 1.6 and 2.9 JND. The three higher distortion levels are above 4 JND and therefore correspond to clearly visible distortions. 
Moreover, the average difference between consecutive  distortion levels in TID2013 is approximately 2 JND, which limits its applicability for benchmarking image quality methods for fine-grained quality differences.

Fig.~\historef{} also shows the quality range for the JPEG AIC-3 dataset. This dataset covers the quality range of approximately 0 to 2.5 JND, with fine-grained differences between distortion levels of about 0.25 JND. However, this dataset includes only five source images and six codecs. In comparison, the proposed AIC2026 dataset covers a larger and more diverse set of 70 source images and 17 codecs  and coding configurations, including  conventional and  learning-based codecs. For each source-codec pair, it includes 20 distortion levels, with an average spacing of approximately 0.2 JND as estimated by CVVDP-based mapping. This makes AIC2026 a strong benchmark dataset for the evaluation of IQA methods for fine-grained quality differences and for assessing artifacts produced by modern codecs, particularly emerging learning-based image compression methods.

\begin{figure*}[ht!]
\centering
\includegraphics[width=0.9\textwidth]{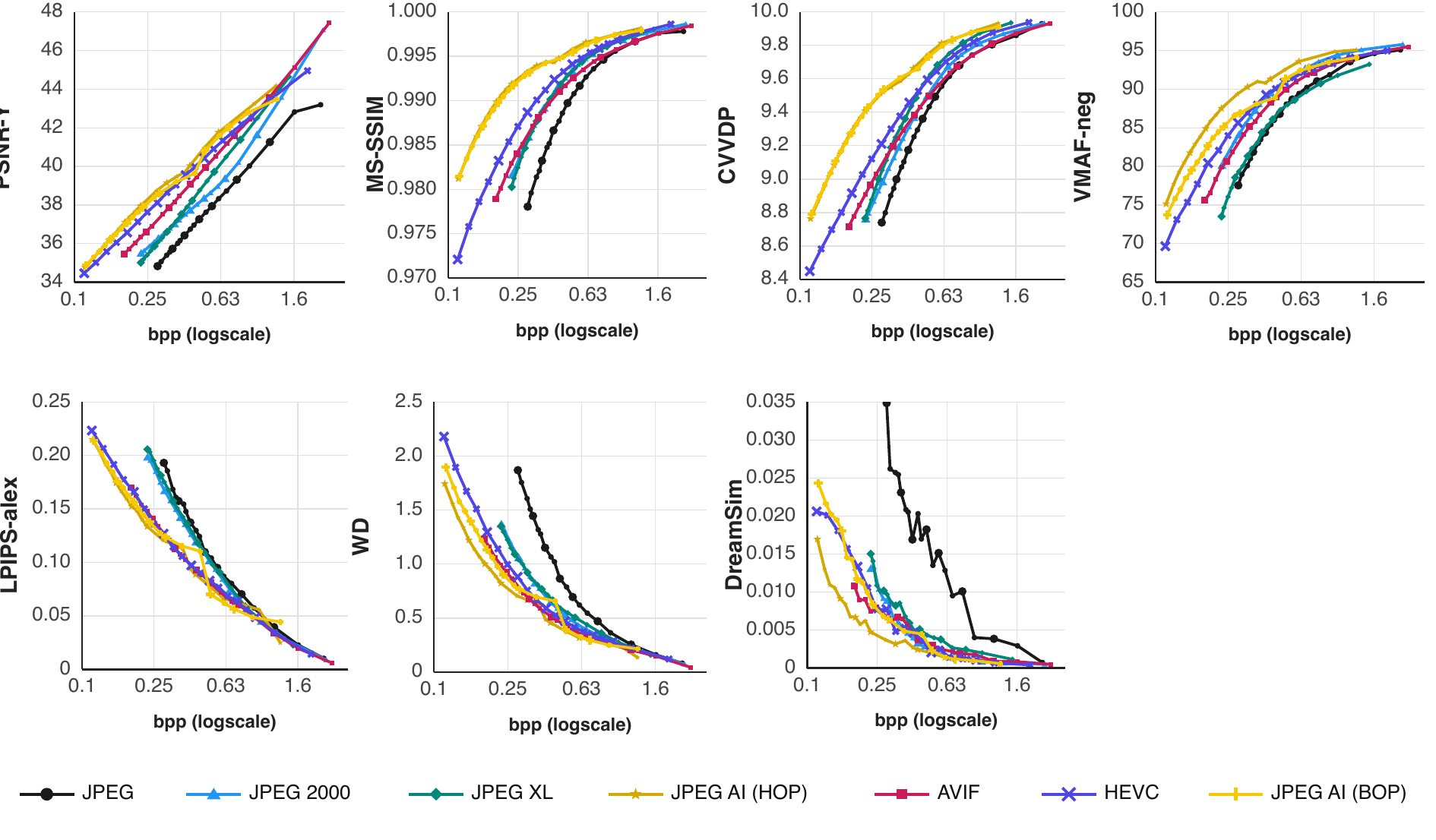}%
\makebox[0pt][r]{%
    \raisebox{2cm}{%
      \includegraphics[width=.15\linewidth]{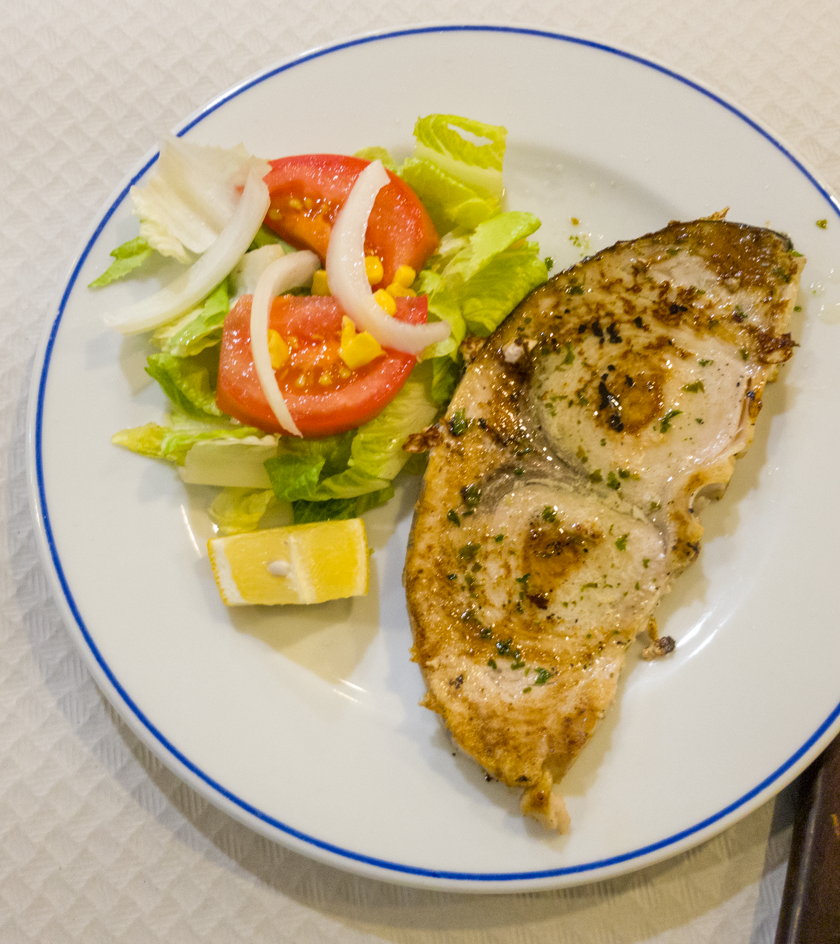}%
    }\hspace*{1em}%
  }%
  
\vspace{-10pt}
\caption{Example distortion-rate curves for source 61, with metric disagreement 0.137.}
\label{fig:example_rd_plots_61}
\end{figure*}

\begin{figure*}[ht!]
\centering
\includegraphics[width=0.9\textwidth]{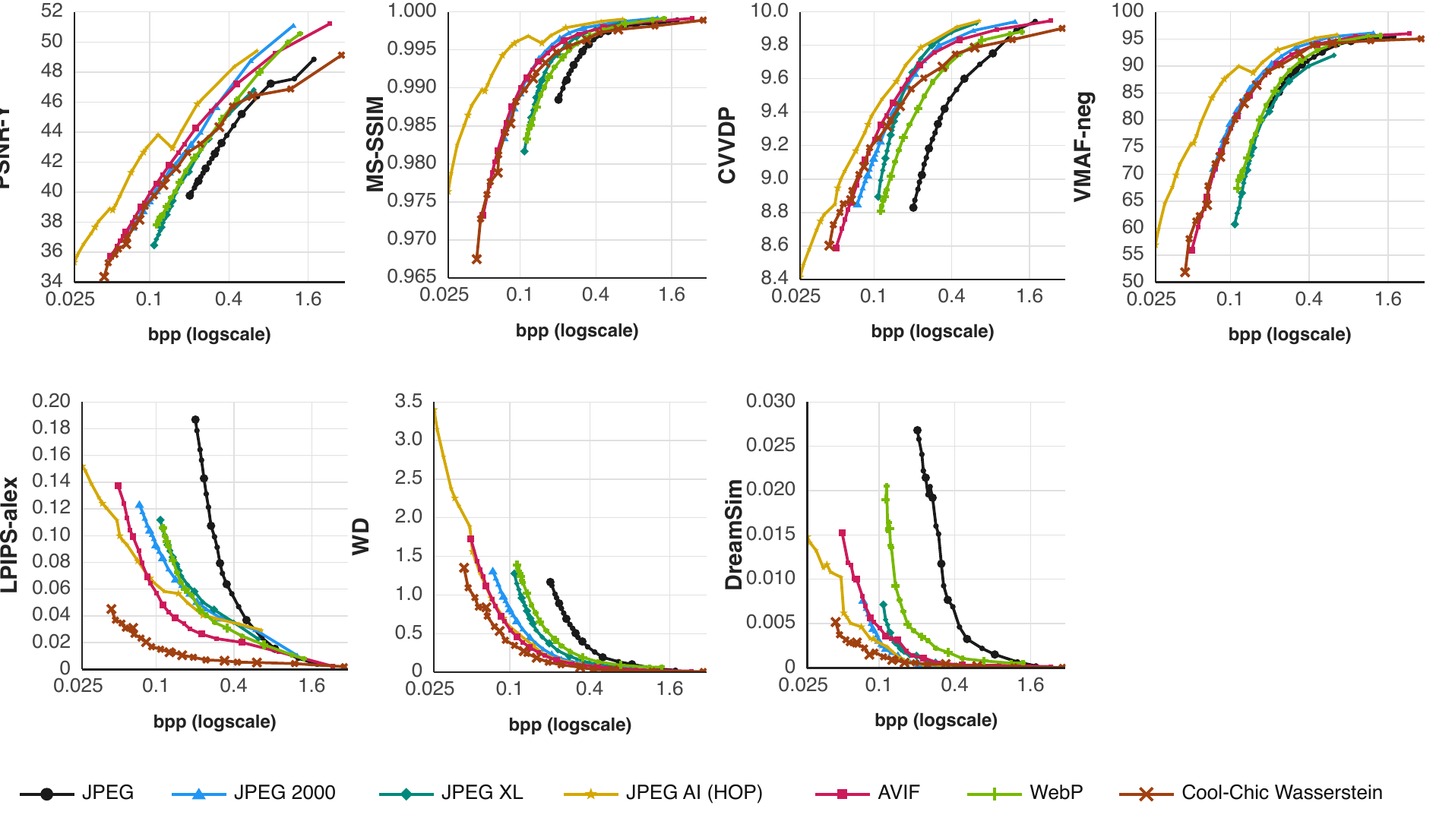}%
\makebox[0pt][r]{%
    \raisebox{2cm}{%
      \includegraphics[width=.15\linewidth]{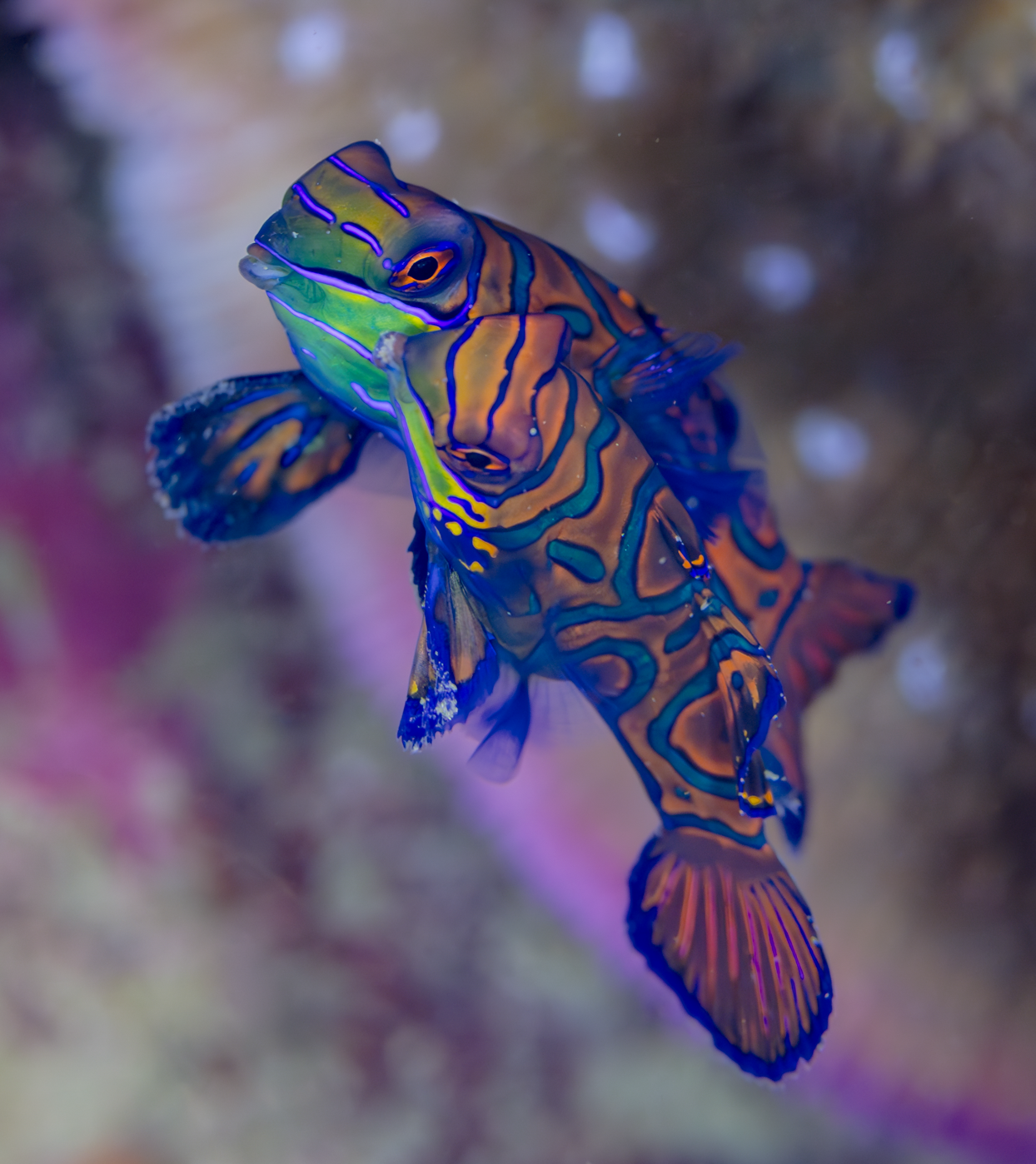}%
    }\hspace*{1em}%
  }%
  
\vspace{-10pt}
\caption{Example distortion-rate curves for source 45, with metric disagreement 0.252.}
\label{fig:example_rd_plots_45}
\vspace{-10pt}
\end{figure*}

The 70 source images were each compressed using 7 codecs, yielding 490 source-codec pairs, each corresponding to a distortion-rate curve.  Fig.~\ref{fig:example_rd_plots_61} shows example distortion-rate curves for source image 61, which has a low disagreement score of 0.137 computed using Equation~\ref{eqn:metric_disagreement} over all pairs of metrics listed in Table~\ref{tab:metrics}. 
The shown distortion-rate curves were estimated using four conventional metrics: PSNR-Y, MS-SSIM \cite{ssim}, CVVDP~\cite{mantiuk2023cvvdp}, and VMAF-neg \cite{vmaf-neg}; and three deep learning-based IQA models: LPIPS~\cite{zhang2018lpips}, WD \cite{qiu2024wd}, and DreamSim~\cite{fu2023dreamsim}. This source image was encoded using the five base codecs: JPEG, JPEG 2000, JPEG XL, AVIF, and JPEG AI (HOP), 
and two codecs from the extended set: HEVC and JPEG AI (BOP).
For this source image, the IQA metrics show relatively high agreement. Across most bitrate points, the compressed versions produce similar metric trends: JPEG reconstructions are generally assigned lower quality, JPEG~AI reconstructions higher quality, and the other reconstructions follow a broadly consistent ordering. Visual inspection of the compressed images shows artifacts including  degraded edges around the plate and onions, and color shifts in the salad. These distortions are ranked similarly by the different IQA metrics.

Fig.~\ref{fig:example_rd_plots_45} shows a contrasting example for source image 45, encoded with the same five base codecs and two additional codecs from the extended set, namely WebP and Cool-Chic Wasserstein. The disagreement score for this image is 0.252. 

Here we see larger discrepancies between the metrics and different IQA metrics give different assessments of the reconstructed images. The Cool-Chic Wasserstein reconstructions receive among the best scores from LPIPS, WD, and DreamSim, but lower scores from several conventional metrics. Differences are also visible among the conventional metrics; for example, in the range above 0.2 bpp, CVVDP and VMAF-neg give different assessments of the JPEG~XL reconstructions. Although different IQA metrics tend to agree on the relative quality of the reconstructed images when quality differences are noticeable, their disagreement increases as the quality differences between distorted images become more subtle.

To facilitate inspection of the distortion-rate curves for all source-codec pairs, we provide an interactive interface
(see Section~\ref{interactive_visualization}).

We evaluated the AIC2026 dataset using 24 conventional and 12 learning-based full-reference IQA metrics, as listed in Table~\ref{tab:metrics}. 
This table reports the pairwise rank correlations among the  IQA metrics on AIC2026, using SRCC in the lower triangular matrix and KRCC in the upper triangular matrix.

The table shows that the evaluated IQA metrics differ substantially in how they rank the compressed images in AIC2026. Some metrics are nearly redundant. For example, PSNR and PSNR-Y are highly correlated, and the two VMAF~v0 variants produce close rankings. A similar pattern is observed within the SSIM family, where SSIM, MS-SSIM, DSSIM, and IW-SSIM show high correlations.  Other metrics show more distinct behavior. PSNR-based measures are less correlated with many perceptual metrics, which suggests that error-based fidelity does not provide the same ordering as metrics that include structural or perceptual modeling. Among the learning-based metrics, DISTS, TOPIQ, and DMM show relatively broad agreement with many of the stronger conventional and perceptual metrics, whereas LPIPS, PieAPP, WaDIQaM, AHIQ, and DeepDC show weaker and more variable correlations. Thus, the main distinction is not simply between conventional and learning-based metrics; rather, the table reveals groups of metrics that behave similarly and others that provide different rankings.

This is important for AIC2026 because the dataset contains compressed images with subtle quality differences, either between adjacent distortion levels or across codecs at the same distortion level. For such fine-grained quality differences, two metrics may agree on the ranking of severely distorted images yet disagree on the ordering of samples separated by small perceptual quality differences. 

\vspace{-10pt}
\subsection*{Granularity and inter-metric disagreement}

To analyze inter-metric disagreement as a function of distortion-level spacing, where the spacing $\delta$ represents the quality difference between consecutive distortion levels, the same number \(n\) of distortion levels is used for every value of $\delta$. This keeps the number of images and pairwise comparisons constant and preserves, for every source, the same balance between same-codec and cross-codec comparisons. Therefore, observed changes in inter-metric disagreement reflect the effect of distortion-level spacing rather than changes in the composition of the evaluated pairs.

Let $k$ denote the spacing in distortion-level indices, corresponding to a CVVDP-based JND spacing of $\delta=0.2k$ between consecutive selected distortion levels. For a given source image $s$, we consider all subsets of $n$ distortion levels whose consecutive elements are separated by $k$ levels. The number of 
such subsets is
$
m=\max\!\left(0,\,20-(n-1)k\right).
$

For values of $n$ and $k$ such that $m>0$, the inter-metric disagreement is defined as
\begin{equation}
\imd_M^{0.2k,n}(s)
=
\frac{1}{m}
\sum_{i=1}^{m}
\imd_M\!\left(
s\big|_{\{\,i+jk \mid 0\leq j<n\,\}}
\right).
\end{equation}
where $s|_D$ denotes the subset of distorted images for the source $s$ whose distortion-level indices belong to the set $D$.
Thus, $\imd_M^{0.2k,n}(s)$ averages the disagreement over all valid subsets with the same number of distortion levels and the same inter-level spacing, thereby reducing dependence on the particular starting distortion level.

\begin{figure}
\includegraphics[width=\linewidth]{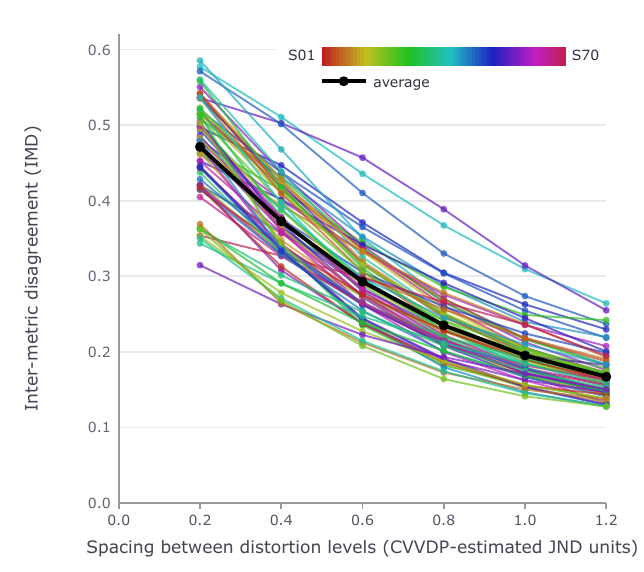}
\caption{Inter-metric disagreement $\imd_M^{\delta,4}(s)$ per source $s$, computed over subsets of $n=4$ of the 20 distortion levels of AIC2026, with different spacing $\delta$ between the levels.} 
\label{fig:imd_granularity}
\end{figure}

Fig.~\ref{fig:imd_granularity} shows the effect of granularity, represented by the JND spacing $\delta$ between consecutive distortion levels, on inter-metric disagreement for $n=4$. Finer distortion-level spacing yields higher inter-metric disagreement, whereas $\imd$ decreases as the spacing becomes coarser. This trend is consistent across all source images.

The fine-grained distortion-level spacing of AIC2026 makes it particularly valuable for benchmarking and developing IQA metrics capable of reliably discriminating the subtle quality differences induced by a wide range of compression artifacts.

\section{Online dataset resources}
\label{sec:online_resources}
To support reproducibility and further research, we make the complete AIC2026 dataset publicly available under  CC BY-SA 4.0 license at \url{https://doi.org/10.18419/DARUS-6156}.

The release includes the complete AIC2026 dataset: 70 source images with metadata, license information and attributions for the dataset and individual images, compressed bitstreams, decoded images for all codecs and operating points, full-resolution distorted images, 840 $\times$ 944 crops to support subjective studies, bitrate and 
objective IQA metric scores.

\label{interactive_visualization}
 An interactive visualization platform at \url{https://pime.uni-konstanz.de/datasets/aic2026/} is also provided. The platform supports visual inspection of source and distorted images, distortion-rate analysis, metric-score distributions, and inter-metric correlation analysis.

\section{Conclusion and future work}
This paper introduces AIC2026, a large-scale dataset for  fine-grained assessment of compressed image quality.  The dataset contains 70 diverse, high-quality source images selected from 2,787 candidates using semantic clustering, inter-metric disagreement analysis, and manual refinement. It provides 9,618 distorted images generated using 17 coding configurations derived from eight conventional codecs and four learning-based codecs. For each source-codec pair, 20 compressed images were selected at approximately perceptually uniformly spaced distortion levels spanning the range 0.2--4.0 JND.

AIC2026 addresses important limitations of existing image quality datasets. It combines large-scale and diverse source content with dense fine-grained perceptual sampling and broad coverage of codec-specific artifacts. To our knowledge, AIC2026 is the first large-scale image compression dataset  that covers fine-grained distortion levels and a wide range of compression  artifacts produced by  conventional codecs and emerging learning-based image compression methods. This combination provides a comprehensive testbed for analyzing distortion-rate behavior and for evaluating whether objective IQA metrics can reliably distinguish subtle quality differences across these diverse compression artifacts.

Our evaluation of 24 conventional and 12 learning-based full-reference IQA metrics revealed substantial disagreement in the quality rankings produced by current methods. The disagreement was especially pronounced for several source-codec pairs involving learning-based compression. These results also show that inter-metric disagreement increases for fine-grained quality differences, both among distorted images from the same source–codec pair with fine-grained spaced perceptual distortion levels and among images from different source–codec combinations at the same distortion levels. Therefore, AIC2026 provides a challenging and valuable benchmark for identifying the strengths and limitations of current IQA metrics and can guide the development of more reliable quality predictors.

The complete dataset is publicly available. 
The repository also includes fixed-size \(840\times944\)-pixel crops of both the source and compressed images. These crops enable side-by-side pairwise comparisons on standard Full HD displays to support crowdsourced subjective studies.

In future work,  a large-scale crowdsourced subjective study following the JPEG AIC-3 methodology will be conducted. The study will estimate a fine-grained perceptual distortion-rate function for each source-codec pair. The resulting subjective scores will provide the perceptual ground truth required to determine which objective metrics most reliably predict subtle quality differences across conventional and learning-based codecs. They will also establish AIC2026 as a benchmark for evaluating metrics submitted to JPEG AIC-4 and for the development of the next generation of objective IQA methods for high-fidelity image compression.

\ifCLASSOPTIONcaptionsoff
\newpage
\fi
\bstctlcite{IEEEexample:BSTcontrol}
\bibliographystyle{IEEEtran}
\bibliography{bstctl,main}

\newpage
\clearpage

\end{document}